\documentclass[10pt]{iopart}
 \pdfoutput=1
\usepackage{epsfig,graphicx,  setspace, mathrsfs}
\usepackage{amssymb,tikz,wrapfig}
\usepackage{txfonts}
\usepackage[titletoc]{appendix} 
\graphicspath{{graphics/}}
\newcommand{\N}{{\rm I\!N}}

\newcommand{\vsp}{\vspace*{3mm}}
\newcommand{\extr}{~{\rm extr}}
\newcommand{\bra}{\langle}
\newcommand{\ket}{\rangle}

\newcommand{\order}{{\cal O}}

\newcommand{\bc}{\mbox{\boldmath $c$}}

\newcommand{\bk}{\mbox{\boldmath $k$}}
\newcommand{\bzeta}{\mbox{\boldmath $\zeta$}}
\newcommand{\bxi}{\mbox{\boldmath $\xi$}}
\newcommand{\bm}{\mbox{\boldmath $m$}}

\newcommand{\bomega}{\mbox{\boldmath $\omega$}}

\newcommand{\bpsi}{\mbox{\boldmath $\psi$}}
\newcommand{\bmu}{\mbox{\boldmath $\mu$}}

\begin{document}
\title{Entropy of random graph ensembles constrained with generalised degrees}
\author{ES Roberts$^{\dag \ddag}$ and ACC Coolen$^{\dag \S}$} 

\address{
${\dag}~$Institute for Mathematical and Molecular Biomedicine, King's College London,  Hodgkin Building,
London SE1 1UL, United Kingdom}
\address{
${\ddag~}$Randall Division of Cell and Molecular Biophysics, King's College London, New
Hunts House, London SE1 1UL, United Kingdom}
\address{$\S~$
London Institute for Mathematical Sciences, 35a South St, Mayfair, London W1K 2XF, United Kingdom}


\ead{ekaterina.roberts@kcl.ac.uk ton.coolen@kcl.ac.uk}

\begin{abstract}
Generalised degrees provide a natural bridge between local and global topological properties of networks. We define the generalised degree $(k,m)$ to be the number of neighbours of a node within one and two steps respectively. 
Tailored random graph ensembles are used to quantify and compare topological properties of networks in a systematic and precise manner, using concepts from information theory. We calculate the Shannon entropy of random graph ensembles constrained with a specified generalised degree distribution. We find that the outcome has a natural connection with the degree-degree correlation which is implied by specifying a generalised degree distribution. We demonstrate how generalised degrees can be used to qualitatively and quantitatively describe a network. 
\end{abstract}

\section{Introduction}


A network is a way of displaying and analysing data. It is interesting to observe how simple local properties can give rise to global topology - for example a critical degree value triggering the inevitable formation of a giant connected component \cite{Dorogovtsev2008Critical}, or a preferential attachment growth model creating a fat tailed degree distribution \cite{Albert2002Statistical}. Understanding these relationships allows insights into the features and origins of real networks. It can also protect against the risk of reporting spurious patterns that are mathematical consequences of already known properties.

We advocate rigorously quantifying topological patterns by viewing them as constraints on a random graph ensemble. This provides a way to measure and compare topological features from the very rational point of view of whether they are present in a large or small number of possible networks. 

A previous paper \cite{Annibale09} considered tailored random graph ensembles with controlled degree distribution and degree-degree correlations; \cite{Roberts11} covered the case of directed networks. In each case, the strategy is to calculate the Shannon entropy. Related quantities such as complexity and information-theoretic distances naturally follow. Subsequent papers were devoted to the numerical generation of graphs \cite{Roberts12} from the proposed ensemble families and applications to molecular biology \cite{Fernandes10}.

In the present paper, we extend this work by calculating, in leading order, the Shannon entropy of random graph ensembles constrained with a specified distribution of generalised degrees. The generalised degree distribution gives the probability of finding a node with any given size of first and second neighbourhood. The Shannon entropy provides a way of quantifying how restrictive the constraint is on the ensemble. This analysis is complemented with a qualitative study of generalised degree distributions.  The results obtained show that general degrees form an interesting bridge between local statistics, global topology and motifs.

\section{Generalised Degrees}
\label{sec:GD_Intro}

\begin{figure}[ht]
\centering
\vspace*{10pt}
\begin{minipage}[b]{0.45\linewidth}
\includegraphics[height = 130pt]{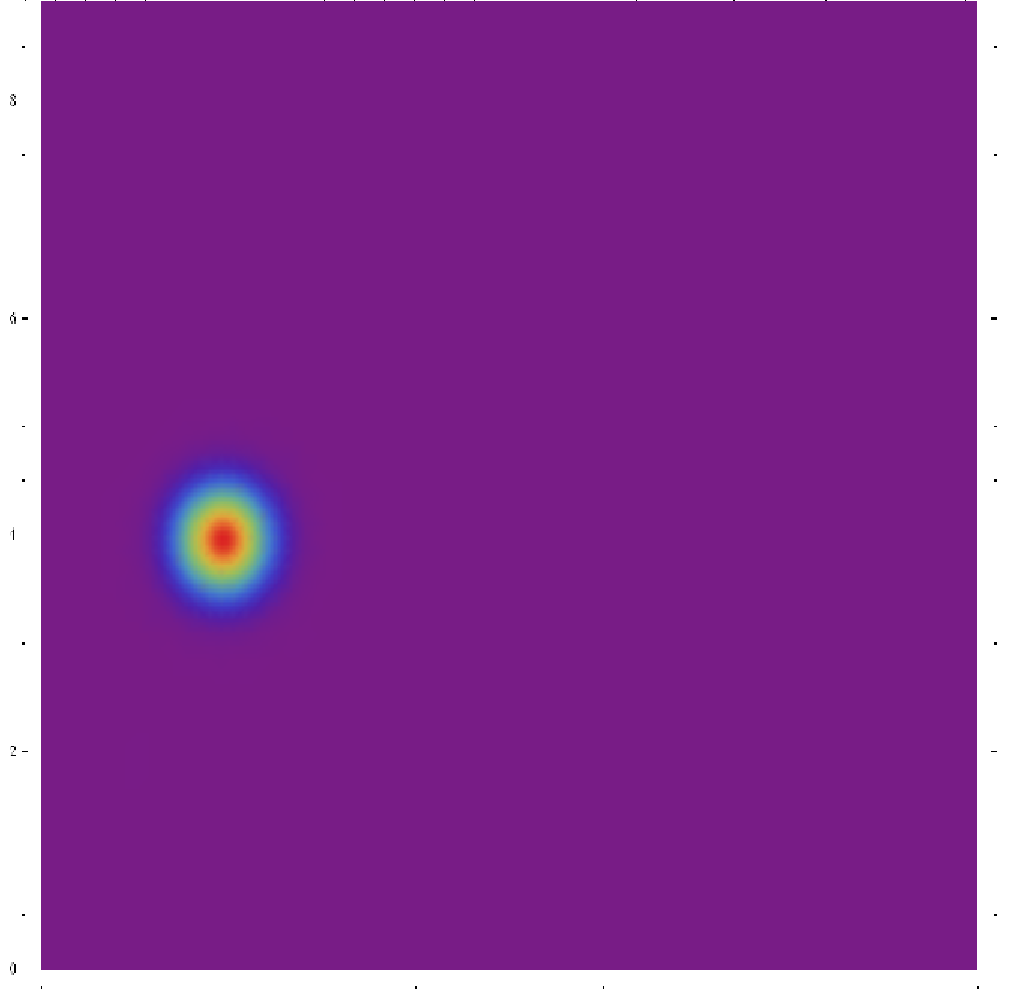}
\end{minipage}
\begin{minipage}[b]{0.45\linewidth}
\vspace{-60pt}
\includegraphics[height = 130pt, angle = 90]{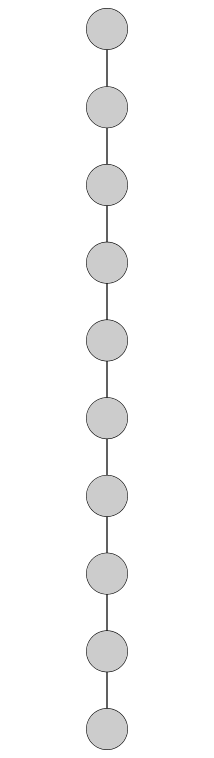}
\end{minipage}
\vspace*{10pt}
\vspace*{10pt}
\begin{minipage}[b]{0.45\linewidth}
\includegraphics[height = 130pt]{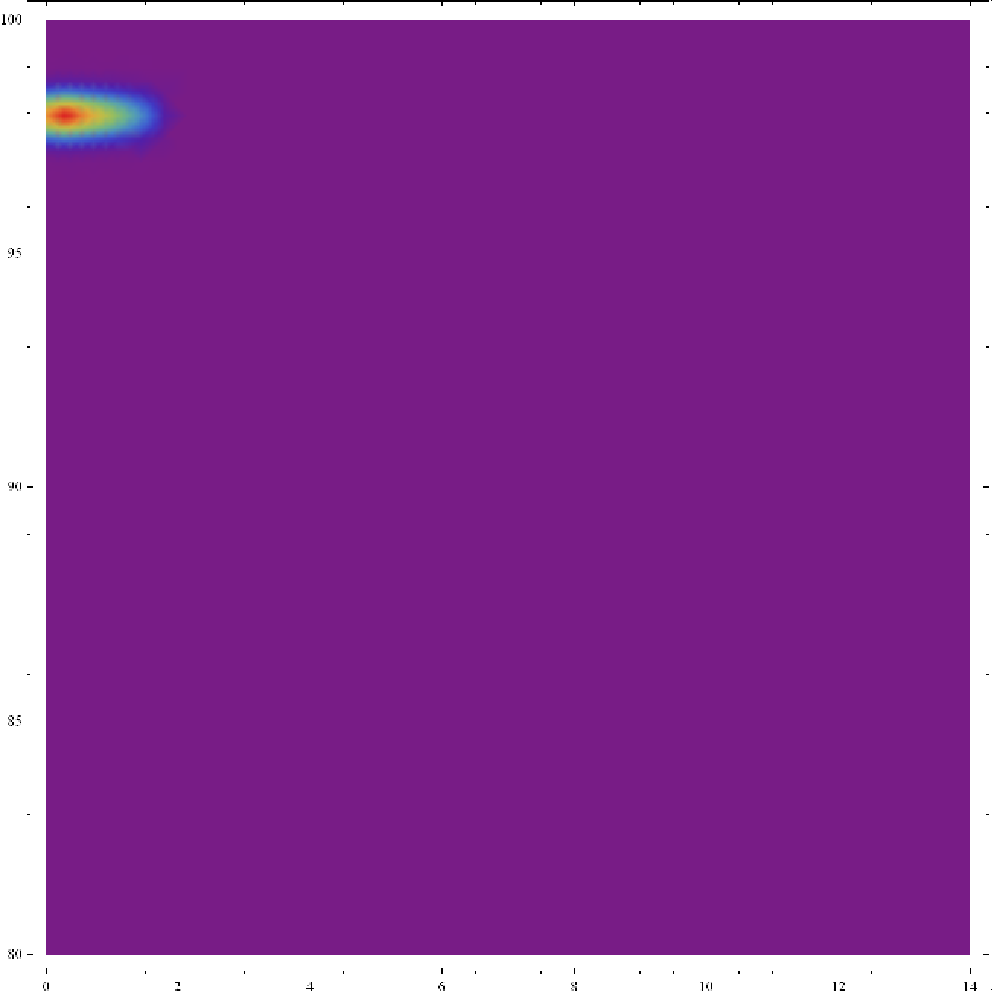}
\end{minipage}
\begin{minipage}[b]{0.45\linewidth}
\includegraphics[height = 130pt]{99_wheel}
\end{minipage}
\vspace*{10pt}
\vspace*{10pt}
\begin{minipage}[b]{0.45\linewidth}
\includegraphics[height = 130pt]{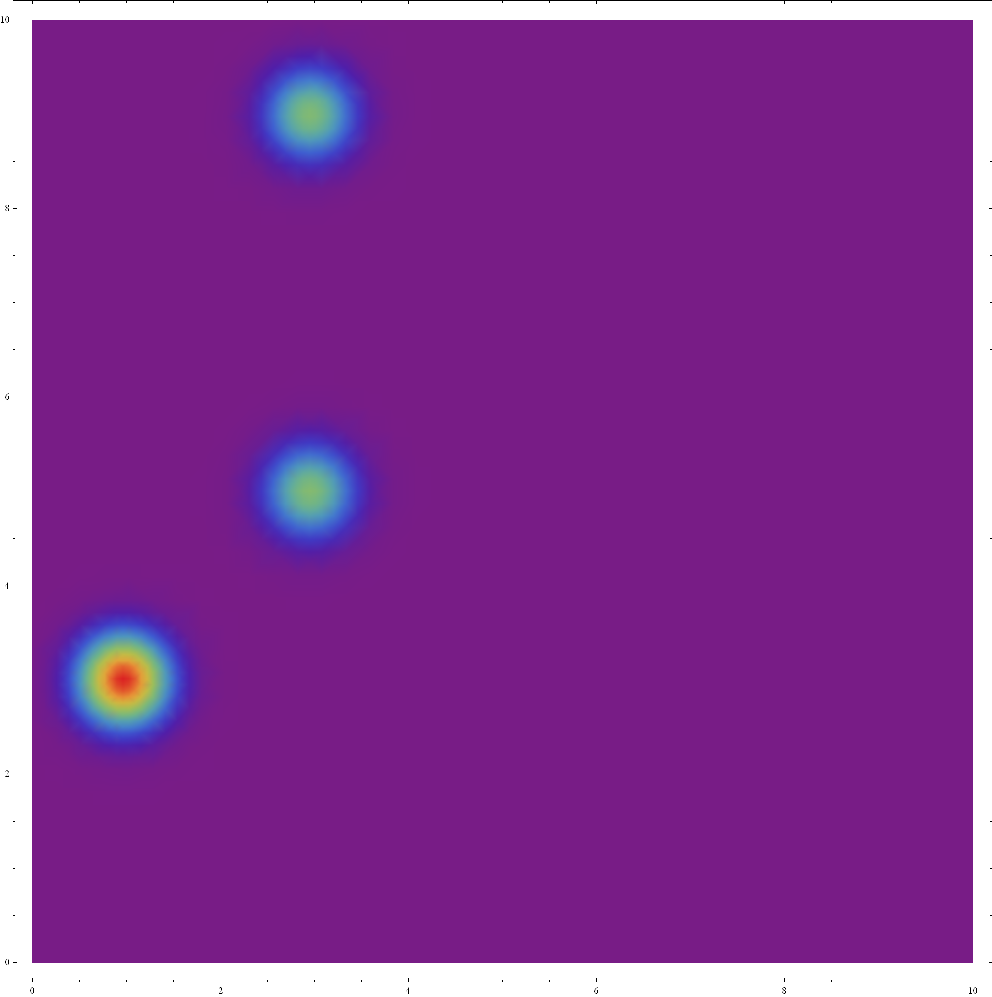}
\end{minipage}
\begin{minipage}[b]{0.45\linewidth}
\includegraphics[height = 135pt]{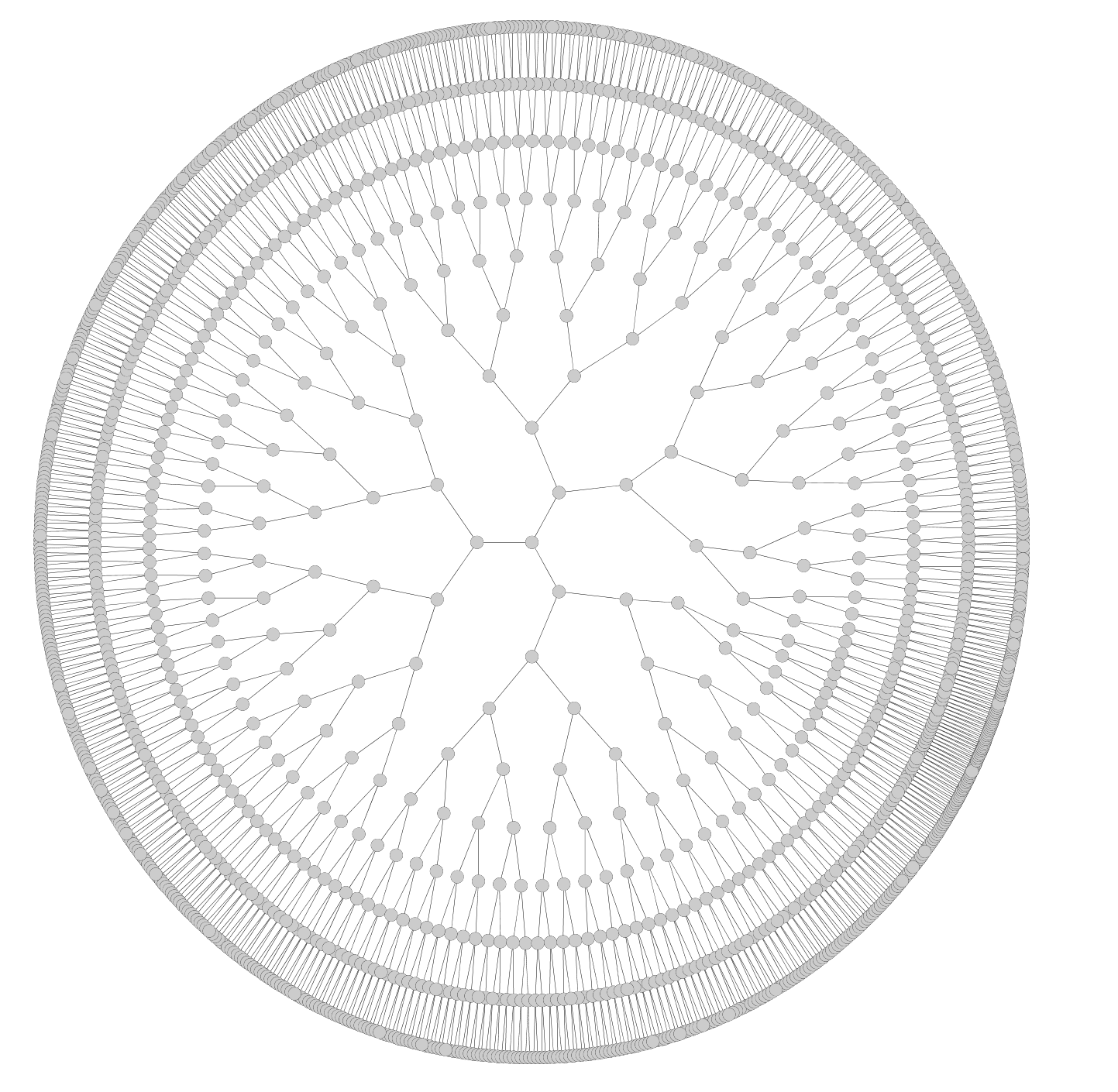} 
\end{minipage}
\vspace*{10pt}
\caption{Some simple networks and their generalised degree heatmap. The first degree is on the $x$ axis and the second degree is on the $y$ axis. The colour at any co-ordinate indicates the frequency of occurrence of that pair of values. The plots are smoothed.

In the first network - a simple chain - every node has the same generalised degree: $(2,4)$. The generalised degree heatmap has a single peak. 

The second network - a star - has generalised degree distribution $p(k,m) = \delta_{(k,m),(99,99)}\frac{1}{100} + \delta_{(k,m),(1,99)}\frac{99}{100}$. There is a visible peak at $(1,99)$. This peak is deformed due to smoothing with the peak at $(99,99)$ (which is too faint to see in its own right).

The final network is a Cayley tree. Every interior node has generalised degree $(3,9)$. View the Cayley tree as being grown one layer at a time (i.e. at every layer attach two new nodes to each node in the preceding layer), for $n$ steps. The final two layers will have irregular generalised degree values (since they lack any further outwards neighbours) - being $(1,3)$ and $(3,6)$ for the final and penultimate layer respectively. Their relative frequencies can be calculated by summing a geometric series, and correspond to the three peaks shown. 
}
\label{fig:synthetic_motif_heatmaps}
\end{figure}

\begin{figure}[ht]
\vsp
\begin{minipage}[b]{0.3\linewidth}
\includegraphics[height = 120pt]{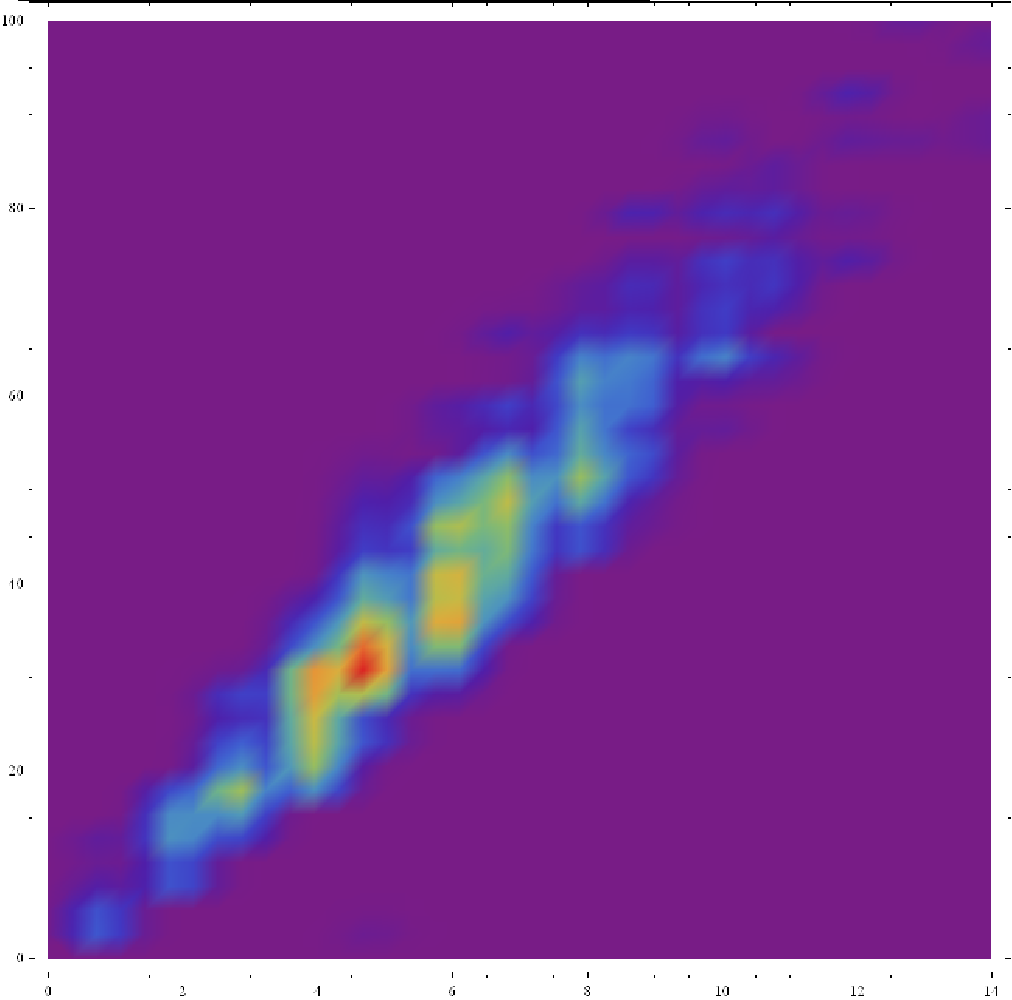}
\end{minipage}\hspace*{4mm}
\begin{minipage}[b]{0.3\linewidth}
\includegraphics[height = 120pt]{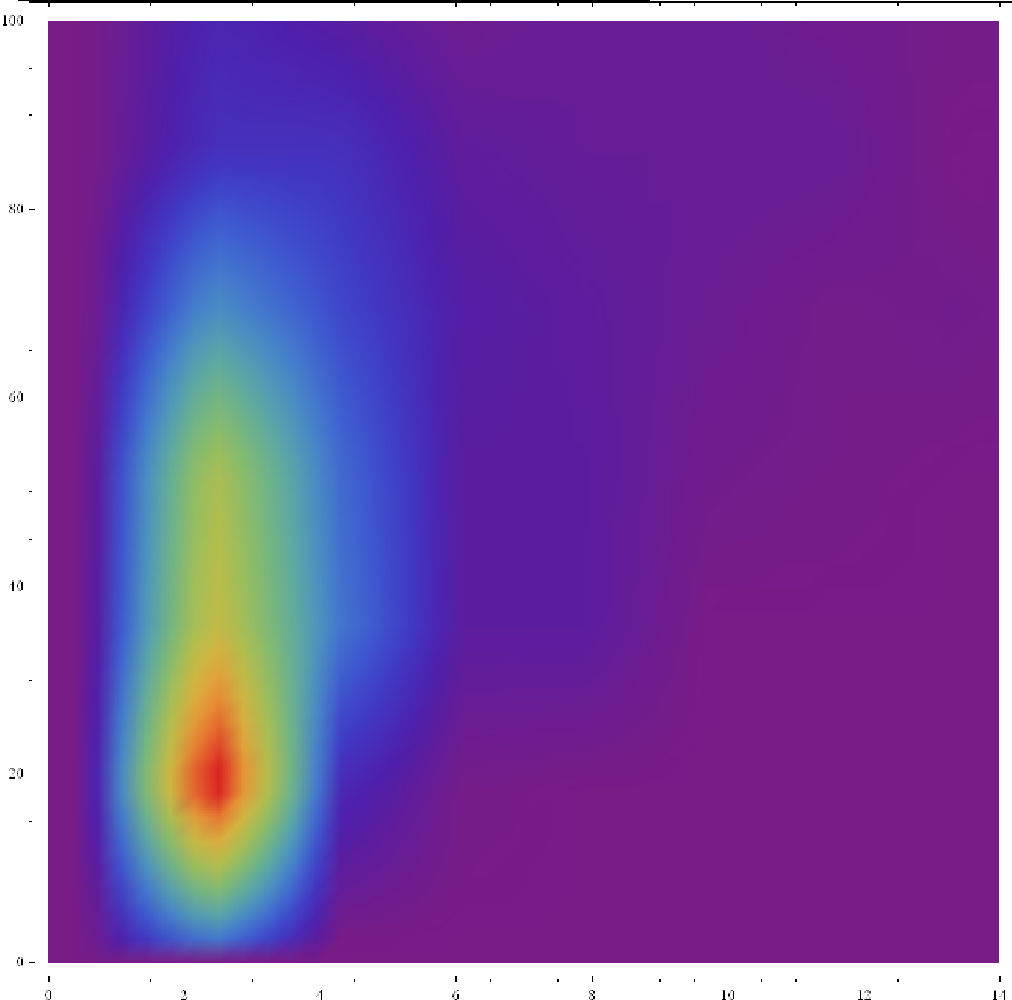}
\end{minipage}\hspace*{4mm}
\begin{minipage}[b]{0.3\linewidth}
\includegraphics[height = 120pt]{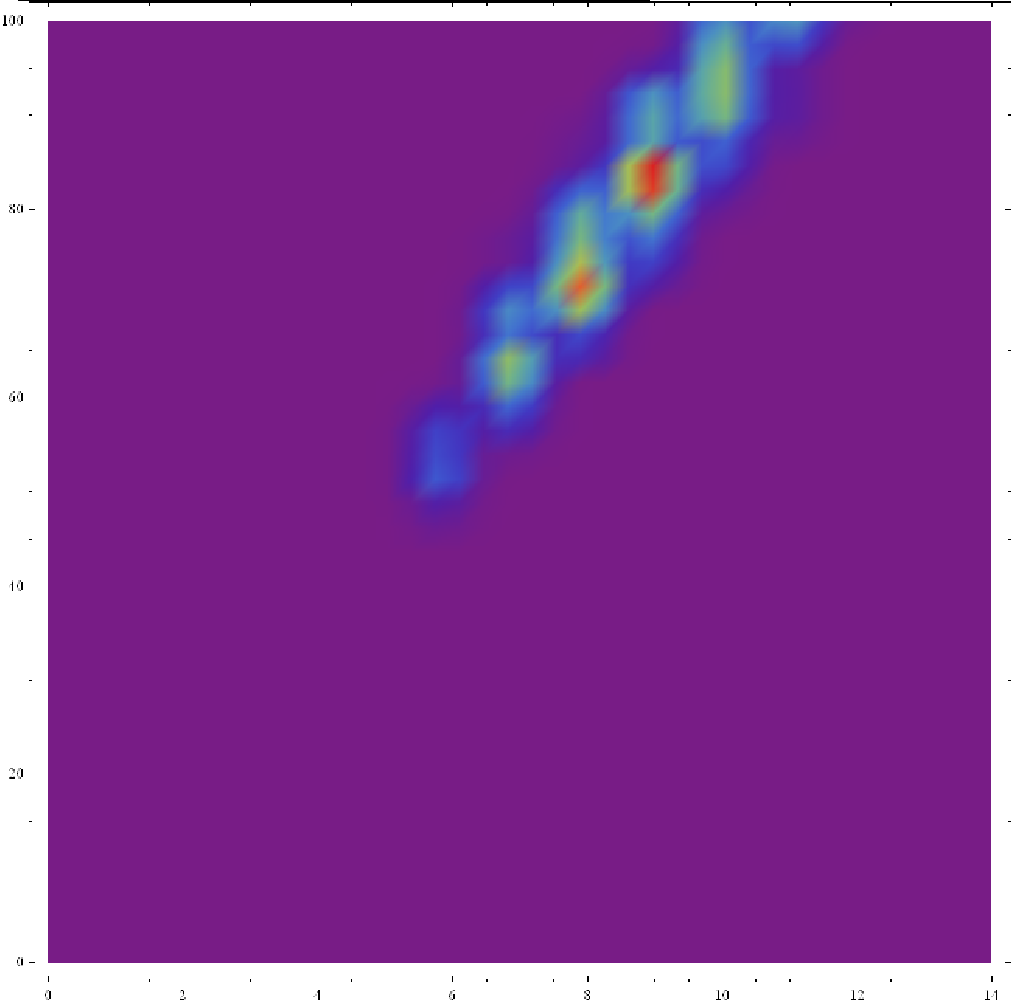}
\end{minipage}
\vsp

\caption{Example generalised degree distributions. From left to right: an Erdos-Renyi type network with links appearing with constant probability \protect\cite{Erdos_Renyi}; a network generated with the Barabasi-Albert preferential attachment model \protect\cite{preferential_attachment}; and, a small-world type random network with enhanced clustering \protect\cite{small_world}. The heatmaps are smoothed; the $x$ axis corresponds to the first degree and the $y$ axis corresponds to the second degree; the intensity of the colour indicates the likelihood of that pair of values. }
\label{fig:erdos}
\end{figure}

Generalised degrees are a natural generalisation of simple degrees. They constitute a much more specific topological signature. Consider Figure \ref{fig:synthetic_motif_heatmaps} which illustrates how certain motifs have characteristic generalised degrees, and Figure \ref{fig:erdos} which shows the generalised degrees of three popular network topologies.

It will be shown in section \ref{sec:W} that the generalised degree has a close connection with the correlations between connected degrees. This links with the concept of the assortativity of the network - whether high degree nodes are likely to be connected to other high degree nodes, or to low degree nodes.

Hence it can be seen that this deceptively simple local constraint imposes strong conditions on the overall topology of the network. The numerical results based on parameters from real biological datasets are given in section \ref{sec:gd_numerical}; applying the formula derived in section \ref{sec:gd_final_answer}  shows that specifying a generalised degree distribution is typically an onerous constraint. 

The premise of the tailored graphs approach is that the most rational way to study a network's overall topology is to view the particular network at hand as a realisation of a particular set of topological constraints. Specifying the generalised degree provides much greater control over the global topology of the networks in the ensemble. 

\subsection{Brief review of literature on generalised degrees}

The concept of a generalised degree was discussed in  \cite{Newmanbook} and references therein. Various relationships were derived using generating functions. 

An alternative definition, presented by the authors of \cite{Faudre}, measured the number of direct neighbours $s$ of a subset of $t$ nodes. The authors of \cite{Faudre} derive conditions based on their definition of general degrees which can ensure that (for some given $m$ and $d$ ) there are at least $m$ internally disjoint paths of length at most $d$. The diameter of the network is an obvious corollary - the smallest $d$ corresponding to $m \geq 1$. These results can be applied to questions of robustness of networks.   

In combinatorics, the generalised degrees concept appears in Seymour's Second Neighbourhood Conjecture, which asserts that in every simple directed graph there exists a vertex $v$ whose second out-neighbourhood contains at least as many vertices as its first out-neighbourhood. 

The authors of \cite{rogers2010spectral} studied the spectral density of random graphs with hierarchically constrained topologies. This includes consideration of generalised degrees, as well as more general community structures. Using the replica method, in a similar way to \cite{bianconi2008entropies}, they achieve a form analogous to equation \eref{eq:intermediate_form}. They proceed numerically from that point, hence our approach to an analytical solution presented in section \ref{sec:gd_final_answer} is entirely novel.

\clearpage

\section{Specifying the problem}
\label{sec:calculation}

\subsection{Definitions}

\label{sec:definitions}
 We consider ensembles of non-directed random graphs.
Each graph is defined by a
symmetric matrix $\bc=\{c_{ij}\}$, with $i,j\in\{1,\ldots,N\}$ and
with $c_{ij}\in\{0,1\}$ for all $(i,j)$. Two nodes $i$ and $j$ are connected by a link if and only if  $c_{ij}=1$. We put $c_{ii}=0$ for all $i$. For simplicity we shall assume that there are no isolated nodes ($p(k_i=0) = 0 $). The first degree, $k_i(\bc)$ and the second degree $m_i(\bc)$, are defined
\begin{eqnarray}
k_i(\bc)=\sum_j c_{ij}~~~~~~~~
\label{eq:basics1}
m_i(\bc)=
\sum_{jk}c_{ij}c_{jk} =
\sum_{j}c_{ij}k_{j}(\bc)
\end{eqnarray}
The first degree $k_i$ counts the number of links to site $i$. The second degree $m_i$ counts the number of distinct two-step paths originating from node $i$. 

We can now define our random graph ensemble. We first draw all pairs $(k_i,m_i)\in\N^2$ for all $i=1\ldots N$ independently at random from a given joint distribution $p(k,m)$, and define the two vectors $\bk=(k_1,\ldots,k_N)$ and $\bm=(m_1,\ldots,m_N)$.
We then generate with uniform probabilities all graphs that meet the topological demands that $\bk(\bc)=\bk$ and $\bm(\bc)=\bm$
\begin{eqnarray}
p(\bc)&=& \sum_{\bk}\sum_{\bm} p(\bk,\bm) \, p(\bc|\bk,\bm)~~~~~~p(\bk,\bm)=\prod_i p(k_i,m_i)
\label{eq:ensemble}
\\ \nonumber
p(\bc|\bk,\bm)&=& Z^{-1}(\bk,\bm) \prod_i\Big(\delta_{k_i,k_i(\bc)}\,  \delta_{m_i,m_i(\bc)}\Big)\\ \nonumber
Z(\bk,\bm)&=&\sum_{\bc}\prod_i\Big(\delta_{k_i,k_i(\bc)}\,  \delta_{m_i,m_i(\bc)}\Big)
\end{eqnarray}
Equivalently
\begin{eqnarray}
p(\bc)&=& Z^{-1}(\bk(\bc),\bm(\bc)) \prod_i p(k_i(\bc),m_i(\bc))
\label{eq:ensemble_compact}
\end{eqnarray}
For this ensemble with prescribed topological features
we wish to calculate, in leading order in the system size $N$, the Shannon entropy per node
\begin{eqnarray}
S&=&-N^{-1}\sum_{\bc}p(\bc)\log p(\bc)
\label{eq:S}
\end{eqnarray}
From this, in turn, follows the effective number ${\cal N}$ of such graphs, via ${\cal N}=\rme^{NS}$.
\subsection{Intermediate answer}
\ref{sec:calculation2} sets out the calculation to the point where the method diverges from that used in \cite{Coolen09}. This takes the calculation to the intermediate form set out below.
\begin{eqnarray}
\label{eq:intermediate_form}
\hspace{-72pt}
S \!= \! \frac{\bar{k}}{2}\left(1 \! + \! \log \left(\frac{N}{\bar{k}} \right)\right) \! -\! \!\sum_k p(k,m) \log \! \left(\! \frac{p(k,m)}{\pi(k)}  \! \right)
\!+\!\! \sum_{(k,m)} p(k,m) \log \left[\! \!\sum_{(\xi_1,.,\xi_k)} \!\prod_{s=1}^k \gamma(k, \xi_s) \delta_{m, \sum^k_{i=1} \xi_i} \right]
\!+ \! \epsilon_N
\end{eqnarray}
 The symbol $\bar{k}$ indicates the average degree; $\pi(k)$ is the Poissonian distribution with the same average degree. The sum in the final term should be read as the sum over all sets of $k$ integers $\xi_1...\xi_k$. The unknown expression $\gamma(\cdot , \cdot)$ complies with the self-consistency relation below.
\begin{eqnarray}
\label{eq:gd_gamma}
\gamma({k, k^\prime}) 
= 
\sum_{m^\prime} \frac{k^\prime}{\bar{k}} \, p({k}^\prime, {m}^\prime) 
\left[
	\frac
	    {
	    	\sum_{\xi^1...\xi^{k^\prime-1}} 
	    		\delta_{m^\prime-k, \sum_{s=1}^{k^\prime-1}\xi^s}
	    		\prod_{s=1}^{k^\prime-1} \gamma\left( k^\prime, \xi^s \right)
	    }
	    {
	    	\sum_{\xi^1...\xi^{k^\prime}} 
	    		\delta_{m^\prime, \sum_{s=1}^{k^\prime}\xi^s}
	    		\prod_{s=1}^{k^\prime} \gamma\left( k^\prime, \xi^s \right)
	    }
\right]
\end{eqnarray}
The self-consistency equation \eref{eq:gd_gamma} does not yield to a straightforward solution. Hence, equation \eref{eq:intermediate_form} is unsatisfying because it can only be evaluated numerically, or in certain special cases. Without a physical interpretation of $\gamma$, this intermediate answer is limited in how much insight it can provide.

%

\section{Implied generalised degree-degree correlation}

\label{sec:W}
Correlations between the generalised degrees of connected nodes can be expressed as
\begin{eqnarray}
W(\vec{k}, \vec{k^\prime}|\bc) &=& \frac{\sum_{ij} c_{ij} \delta_{\vec{k}, \vec{k}_i} 
\delta_{\vec{k}^\prime, \vec{k}_j}
}
{
\sum_{ij} c_{ij}
}
\end{eqnarray}
where $\vec{k} = (k,m)$. We can calculate the average of this quantity in our ensemble with a specified generalised degree distribution.
\begin{eqnarray}
W(\vec{k}, \vec{k^\prime}) &=& \lim_{N \rightarrow \infty} \sum_{\bc} p(\bc|p(\vec{k}))W(\vec{k}, \vec{k^\prime}| \bc)
\end{eqnarray}
where $$p(\bc|p(\vec{k})) = \sum_{\{ \vec{k}_1...\vec{k}_N \}} 
\frac{ \prod_i \delta_{\vec{k}_i, \vec{k}_i(\bc)} }
{\sum_{\bc^\prime} \prod_i \delta_{\vec{k}_i, \vec{k}_i(\bc^\prime)} }
\prod_i p(\vec{k}_i)$$
Expand this expression
\begin{eqnarray}
W(\vec{k}, \vec{k^\prime})
&=&
\frac{1}{\bar{k}N}
\sum_{rs} 
\sum_{\{ \vec{k}_1...\vec{k}_N \}}
\left[
\prod_i 
p(\vec{k}_i)
\delta_{\vec{k}, \vec{k}_r} 
\delta_{\vec{k}^\prime, \vec{k}_s} 
\right]
\frac{ \sum_{\bc} c_{rs}
\prod_i \delta_{\vec{k}_i, \vec{k}_i(\bc)} }
{\sum_{\bc^\prime} \prod_i \delta_{\vec{k}_i, \vec{k}_i(\bc^\prime)} }
\end{eqnarray}
\ref{sec:W_appendix} sets out in detail the key steps of the calculation. In leading order, the integral evaluates to 
\begin{eqnarray}
W(\vec{k}, \vec{k^\prime})
&=&
\frac{k k^\prime}{(\bar{k}N)^2}
\sum_{rs} 
\sum_{\{ \vec{k}_1...\vec{k}_N \}}
\left[
\prod_i p(\vec{k}_i)
\delta_{\vec{k}, \vec{k}_r} 
\delta_{\vec{k}^\prime, \vec{k}_s} 
\right]
\gamma^\star(\vec{k}, k^\prime)
\gamma^\star(\vec{k^\prime}, k)
\end{eqnarray}
for 
\begin{eqnarray}
\gamma^\star(\vec{k^\prime}, k) = \frac
{
\sum_{\{\xi_1...\xi_{k^\prime-1} \}} \prod_{s=1}^{k^\prime-1} \gamma(k^\prime, \xi_s) \delta_{m^\prime-k, \sum \xi_s}}
{\sum_{\{\xi_1...\xi_{k^\prime} \} } \prod_{s=1}^{k^\prime} \gamma(k^\prime, \xi_s) \delta_{m^\prime, \sum \xi_s} }
\end{eqnarray}
satisfying 
\begin{eqnarray}\sum_{m^\prime} \frac{k^\prime}{\bar{k}} p(\vec{k}^\prime)\gamma^\star(\vec{k^\prime}, k) = \gamma(k, k^\prime)
\end{eqnarray}
whereupon we can simplify
\begin{eqnarray}
\label{eq:W}
W(\vec{k}, \vec{k^\prime})
&=& 
\gamma^\star(\vec{k}, k^\prime)
\gamma^\star(\vec{k^\prime}, k)
\frac{k k^\prime p(\vec{k}) p(\vec{k}^\prime)}{\bar{k}^2}
\end{eqnarray}
We can very easily take the marginal of the object $W$ using the relation for $\gamma^\star$
\begin{eqnarray}
\sum_{m, m^\prime} W(\vec{k}, \vec{k}^\prime)&= W({k}, {k^\prime}) =\gamma(k, k^\prime) \gamma(k^\prime, k)
\end{eqnarray}
This relationship is elegant and, at first glance, surprising. Considering this further, however, it becomes clear that a specified generalised degree distribution will essentially induce a non-trivial degree-degree correlation function that - in the limit - will indeed be shared by all members of the ensemble. It would be possible to write down various combinatorial relationships - and it is interesting and encouraging that the statistical mechanics route naturally discovers the same links. 

Section \ref{sec:gd_final_answer} will use equation \eref{eq:W} to eliminate the undetermined $\gamma$ term from equation \eref{eq:intermediate_form}. An interesting alternative route would have been to use equation \eref{eq:W} in order to describe $\gamma$ in terms of topological observables. The form of the self-consistency equation \eref{eq:gd_gamma} is strongly suggestive of a natural interpretation of $\gamma$ in terms of conditional probabilities associated with links in the network.

\section{Completing the solution}
\label{sec:gd_final_answer}

The aim is to resolve equation \eref{eq:intermediate_form} into observable quantities. Consider the term 
\begin{eqnarray}
 \Gamma =  \sum_{(k,m)} p(k,m) \log \left[ \sum_{\{\xi_1, ..., \xi_k \}} \prod_{\xi_1, ..., \xi_k} \gamma(k, \xi) \delta_{m, \sum^k_{i=1} \xi_i} \right]
 \end{eqnarray}
Observe that at this point  of the calculation, the effect of factorising across nodes has been to break the expression down into terms which, for every generalised degree $(k,m)$, enumerate all the possible ways of dividing $m$ second neighbours between $k$ first neighbours. The term inside the logarithm sums over all configurations $\{k: \xi_1... \xi_k \}$ which meet the condition $\sum_{s=1}^k \xi_s = m$.

To formalise this idea, re-aggregate the expression to write
\begin{eqnarray}
 \Gamma =  \frac{1}{N}\log \Bigg\lbrace \prod_{(k,m)} \left[ \sum_{\{\xi_1, ..., \xi_k \}} \prod_{\xi_1, ..., \xi_k} \gamma(k, \xi) \delta_{m, \sum^k_{i=1} \xi_i} \right]^{N p(k,m)} \Bigg\rbrace
 \end{eqnarray}
observing that $Np(k,m)$ can be assumed to be an integer by construction, since we would typically have taken the statistics from a real network. 
This multiplies out to
\begin{eqnarray}
 \Gamma =  \frac{1}{N}\log  \left [ \sum_{\{\xi^1_1, ..., \xi_{k_1}^1 \}}  \sum_{\{\xi^2_1, ..., \xi_{k_2}^2 \}} ....  \sum_{\{\xi^N_1, ..., \xi^N_{k_N} \}} \prod_{\{\mathbf{\xi} \}} \gamma(k_s, \xi_s^t) \delta_{m_s, \sum^{k_s}_{t=1} \xi^t_s} \right]
 \end{eqnarray}
where the frequency with which any pair $(k,m)$ appears is given by $N p(k,m)$

This can be restated as
\begin{eqnarray}
\label{eq:Gamma_plus_Gamma}
\Gamma &= &  \frac{1}{N} \log \left[ \Gamma_1 + \Gamma_2 + \ldots + \Gamma_X \right] 
\end{eqnarray}
where each of the $\Gamma_\star$ is a product of $\bar{k}N$ terms of type $\gamma$. A typical term would be 
\begin{eqnarray} 
\hspace{-70pt}
\label{eq:long_multipliedout_gamma_term}
\gamma(k_1, \xi_1)\gamma(k_1, \xi_2)...\gamma(k_1, \xi_{k_1})\gamma(k_2, \xi^\prime_1)\gamma(k_2, \xi^\prime_2)...\gamma(k_N, \xi^{\prime...\prime}_1)\gamma(k_N, \xi^{\prime...\prime}_2)...\gamma(k_N, \xi^{\prime...\prime}_{k_N})
\end{eqnarray}
We can now see that the separate terms $\Gamma_1, \Gamma_2, \ldots ,\Gamma_X$ precisely enumerate all the permutations of degrees and neighbour-degrees for networks with a generalised degree sequence consistent with any pair $(k,m)$ appearing $N p(k,m)$ times. 

If we identify the $\xi$ as actual degrees and $\gamma(k, k^\prime)$ with an actual edge in the network, then a necessary and sufficient condition for graphicality can be immediately deduced: in equation \eref{eq:long_multipliedout_gamma_term} every $\gamma(k, k^\prime)$ must have a matching $\gamma(k^\prime, k)$. This can be seen by construction. Every edge in the network going from $k_{i_1} \rightarrow k_{i_2}$ will give a contribution of $\gamma(k_{i_1},  k_{i_2})$ when counted in the $j=i_1$ term, and a contribution of $\gamma(k_{i_2},  k_{i_1})$ when counted in the $j=i_2$ term. If this can be done, then it defines a network. If it cannot be done, then it proves that no network can be drawn with that sequence of degree and neighbour-degrees. 

This insight allows the expression to be substantially simplified, since we already know that $\gamma(k, k^\prime)\gamma(k^\prime, k)= W(k^\prime, k)$ where $W(k^\prime, k)$ is the correlation between degrees of connected nodes. Hence, we now claim that a typical term will look like 
$$W(a,b)W(c,d)W(e,f)W(g,h)...W(x,y)W(m,n)W(s,t) $$
where a network is graphical. We disregard non-graphical terms. This step is in keeping with the natural interpretation of the problem, but not rigorously proved to be valid. 

We can use our knowledge of the average quantities in the ensemble in order to deduce that for any arbitrary $\alpha$ and $\beta$ the term $W(\alpha, \beta)$ will appears in this sequence $\bar{k}N W(\alpha, \beta)$ times. 

The number of $\Gamma$ terms is equation \eref{eq:Gamma_plus_Gamma} is $ \mathscr{P}$. This could be expected to be equal to $ \mathscr{P} = \prod_{(k,m)} \left[ \mathscr{P}_{k,m} \right]^{N p(k,m)}$ 
where $\mathscr{P}_{k,m}$ 
is equal to the number of partitions of $m$ into $k$ terms. 
However, if the probability density function does not have compact support, then some of the $\xi_i$ correspond to degrees which have zero probability - and so $\gamma(\star,\xi_i) \equiv 0$. In this case, the number of partitions is more accurately written as $\mathscr{P}_{k,m} = \sum_{\{ \xi_1...\xi_k \}} \delta_{\sum \xi_i , m} \prod_i \left( 1 - \delta_{p(\xi_i),0} \right)$. This enumerates the number of $\Gamma$ terms in \ref{eq:Gamma_plus_Gamma}.  However, some of these sequences will not be graphical, so this is an upper bound. 

So we have reached 
\begin{eqnarray}
\label{eq:answer_gd}
 \Gamma &=& \frac{1}{N}\log \left[ \mathscr{P} \prod_{k, k^\prime} W(k, k^\prime)^{\frac{\bar{k}N}{2} W(k, k^\prime)} \right] \\ \nonumber
 \Gamma &=& \frac{\bar{k}}{2} \sum_{k, k^\prime} W(k, k^\prime) \log W(k, k^\prime) + \sum_{k,m} p(k,m) \log \mathscr{P}_{k,m}
\end{eqnarray}
where the factor of $1/2$ compensates for over counting. An upper bound can be provided for the final term: $\mathscr{P}_{k,m}\leq \sum_{\{ \xi_1...\xi_k \}} \delta_{\sum \xi_i , m} \prod_i \left( 1 - \delta_{p(\xi_i),0} \right)$, or $\mathscr{P}_{k,m}\leq {m-1 \choose k-1}$ if $p(\xi) >0$ for all $ \xi \leq m-k+1$.

Accordingly, the full expression for the entropy per node will be
\begin{eqnarray}
\label{eq:entropy_final}
S &=& \frac{\bar{k}}{2} \left(1 + \log \left(\frac{N}{\bar{k}} \right) \right) - \sum_k p(k,m) \log \left(\frac{p(k,m)}{\pi(k)} \right)  
\\ \nonumber && +
 \frac{\bar{k}}{2} \sum_{k, k^\prime} W(k, k^\prime) \log W(k, k^\prime) + \sum_{k,m} p(k,m) \log \mathscr{P}_{k,m}
+ \epsilon_N\\ \nonumber
\end{eqnarray}

\section{Validation}
\label{sec:gd_synthetic_examples_text}

The restrictiveness of the generalised degree constraint means that it is easy to find examples where the Shannon entropy can be obtained directly. This is useful to validate equation \eref{eq:entropy_final}. Additionally, where the self consistency equation \eref{eq:gd_gamma} can be solved, this can be checked against the final terms of equation \eref{eq:entropy_final}. \ref{sec:gd_synthetic_examples} provides several such examples. Figures \ref{fig:ladder}, \ref{fig:wheel}, \ref{fig:gd_W_example} and \ref{fig:gd_molecule} illustrate the validation examples. 

\subsection{Comparison with the maximum entropy ensemble specified by simple degrees only}

The flat distribution corresponds to $p(m|k) = \sum_{_m\left\lbrace \xi \right\rbrace^k} \prod_i \frac{\xi_i p(\xi_i)}{\bar{k}}$ where $_m\left\lbrace \xi \right\rbrace^k$ describes a set of partitions of $m$ into $k$. This case will also have factorising $W$. 

It can be shown (see e.g. \cite{Newmanbook} )  that the flat ensemble defined by the degree distribution only is given by
\begin{eqnarray}
p_{flat}(k,m) &=& \frac{p(k)}{\bar{k}^k}\sum_{_{m}\{\xi\}^k} \prod_{s=1}^k p(\xi_s)\xi_s
\end{eqnarray}
For the expression in equation \eref{eq:intermediate_form} to be consistent with earlier results, it must collapse back to the degree-only case - as derived in \cite{Annibale09} - for $p(k,m) = p_{flat}(k,m)$. 

To validate the intermediate result given by equation \eref{eq:intermediate_form},  the final form achieved for generalised degrees less the final form achieved for simple degrees must leave a remainder satisfying
\begin{eqnarray}
\sum_{k,m}p(k,m) \log \Gamma(k,m) - \sum_{k,m}p(k,m) \log p(m|k) = 0
\end{eqnarray}
where
\begin{eqnarray}
\Gamma(k,m) = \sum_{(\xi_1, ..., \xi_k)} \prod_{\xi_1, ..., \xi_k} \gamma(k, \xi) \delta_{m, \sum^k_{i=1} \xi_i} 
\end{eqnarray}
Substitute in the result above $p(m|k)= \sum_{_{m-k}\{\xi\}^k} \prod_{s=1}^k \frac{p(\xi_s)\xi_s}{\bar{k}}$. This implies $\gamma(k, \xi_s) = \frac{\xi_s p(\xi_s)}{\bar{k}}$. This satisfies the self-consistency relationship \eref{eq:gd_gamma}.

Since the connected node degrees are independent in this case, it is valid to write
$$ \sum_{k,m} p(k,m)\log \left(\sum_{_m\left\lbrace \xi \right\rbrace^k} \prod_i \frac{\xi_i p(\xi_i)}{\bar{k}} \right) = \sum_{k,m} p(k,m)\log \mathscr{P}_{km}  + \sum_{\xi} \xi p(\xi)\log\left(\frac{\xi p(\xi)}{\bar{k}}\right) $$
which is consistent with the final result in equation \eref{eq:entropy_final}.

\section{Numerical results}
\label{sec:gd_numerical}

\begin{figure}\vsp
\begin{minipage}[b]{0.3\linewidth}
\includegraphics[width = \linewidth]{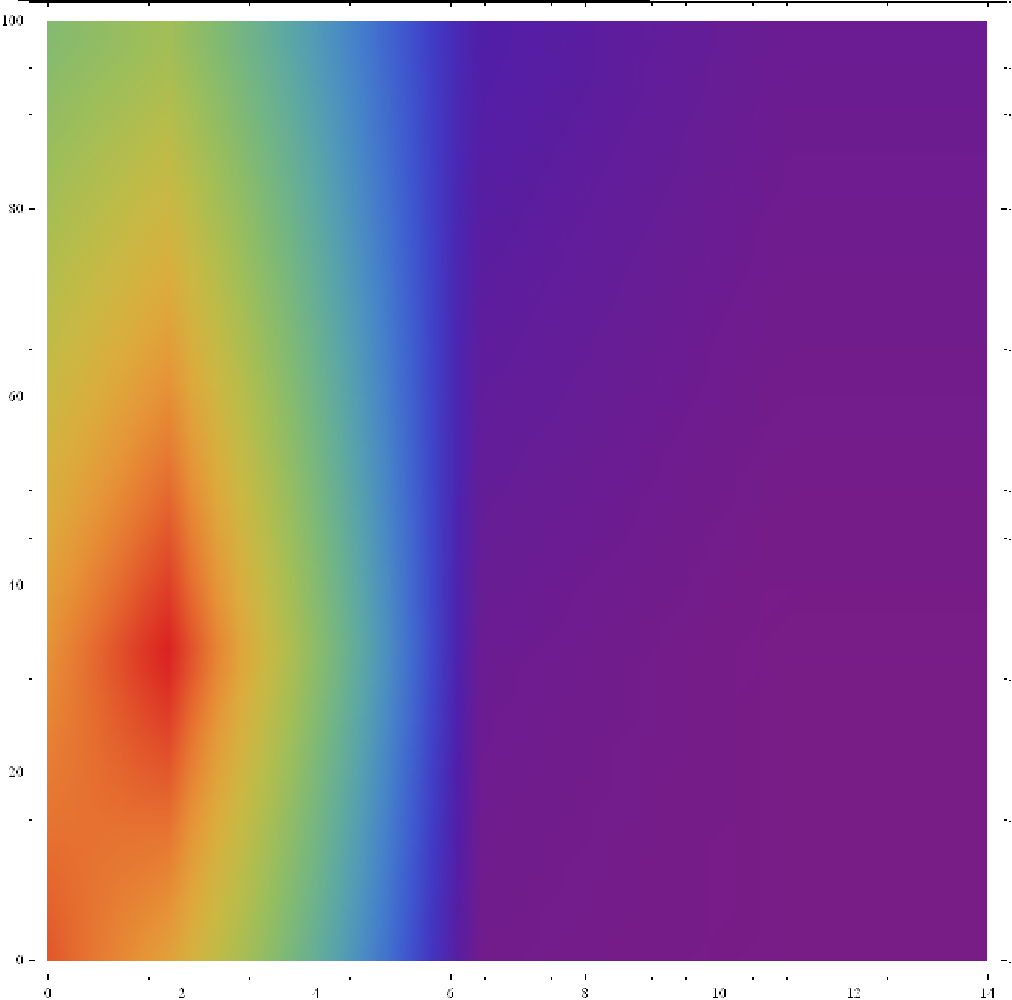}
{\textit{t. pallidum} \protect\cite{tre}}
\end{minipage}
\begin{minipage}[b]{0.3\linewidth}
\includegraphics[width = \linewidth]{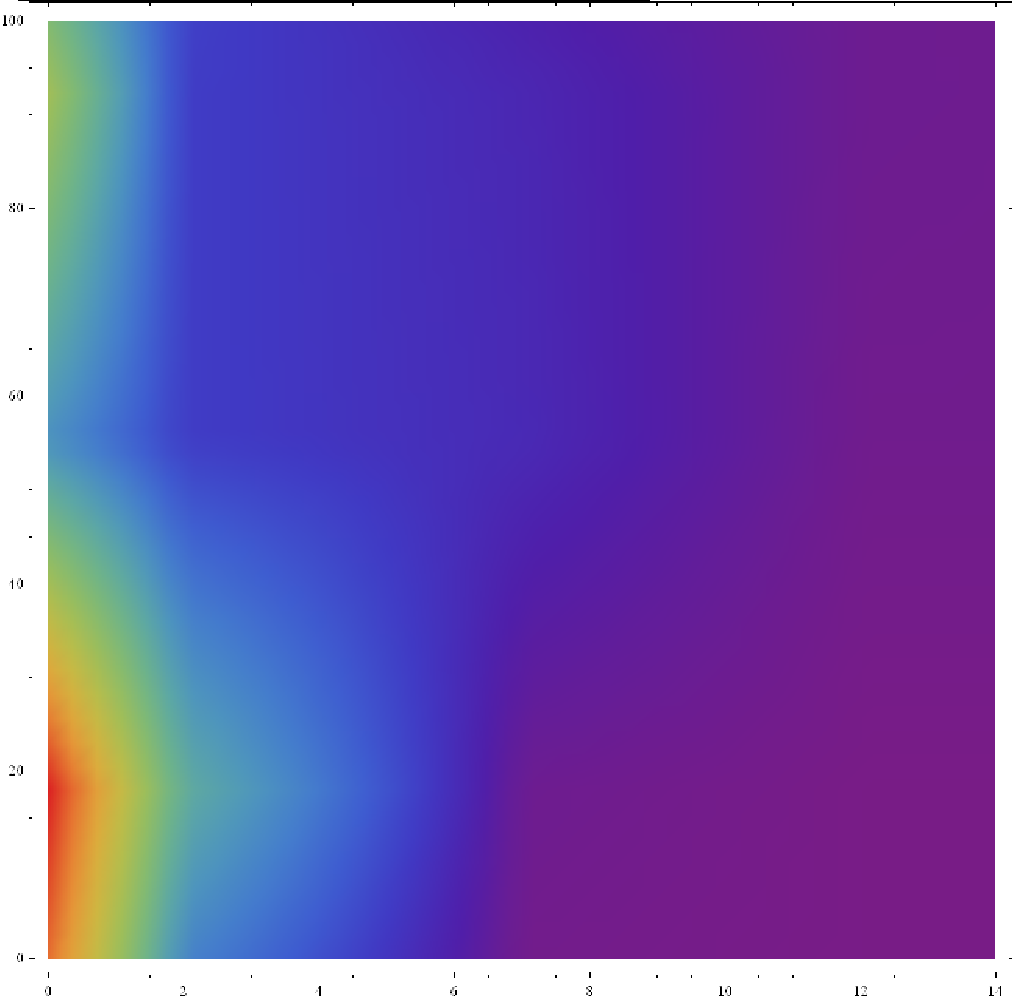}
{\textit{h. sapiens} \protect\cite{ewing}}
\end{minipage}
\begin{minipage}[b]{0.3\linewidth}
\includegraphics[width = \linewidth]{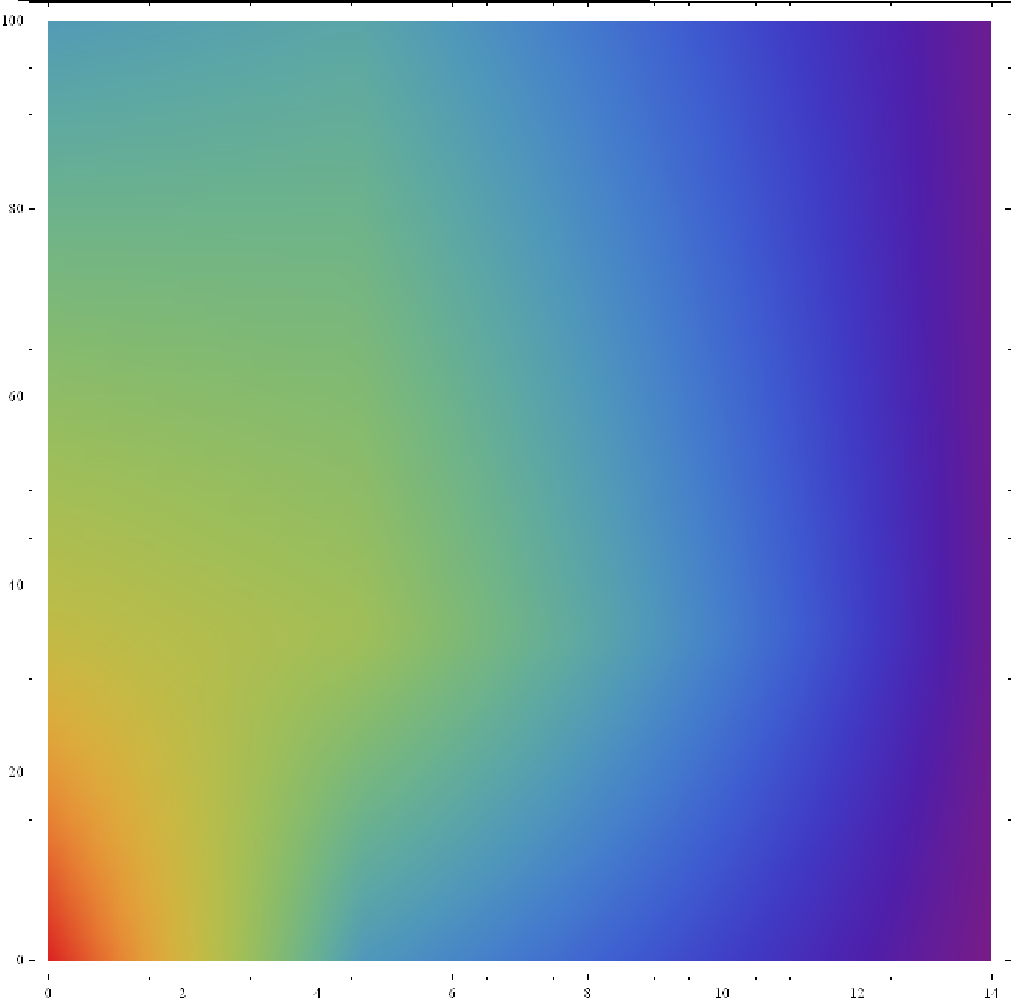}
{\textit{e. coli} \protect\cite{arifuz}}
\end{minipage} 
\newline\vspace*{1mm}

\begin{minipage}[b]{0.3\linewidth}
\includegraphics[width = \linewidth]{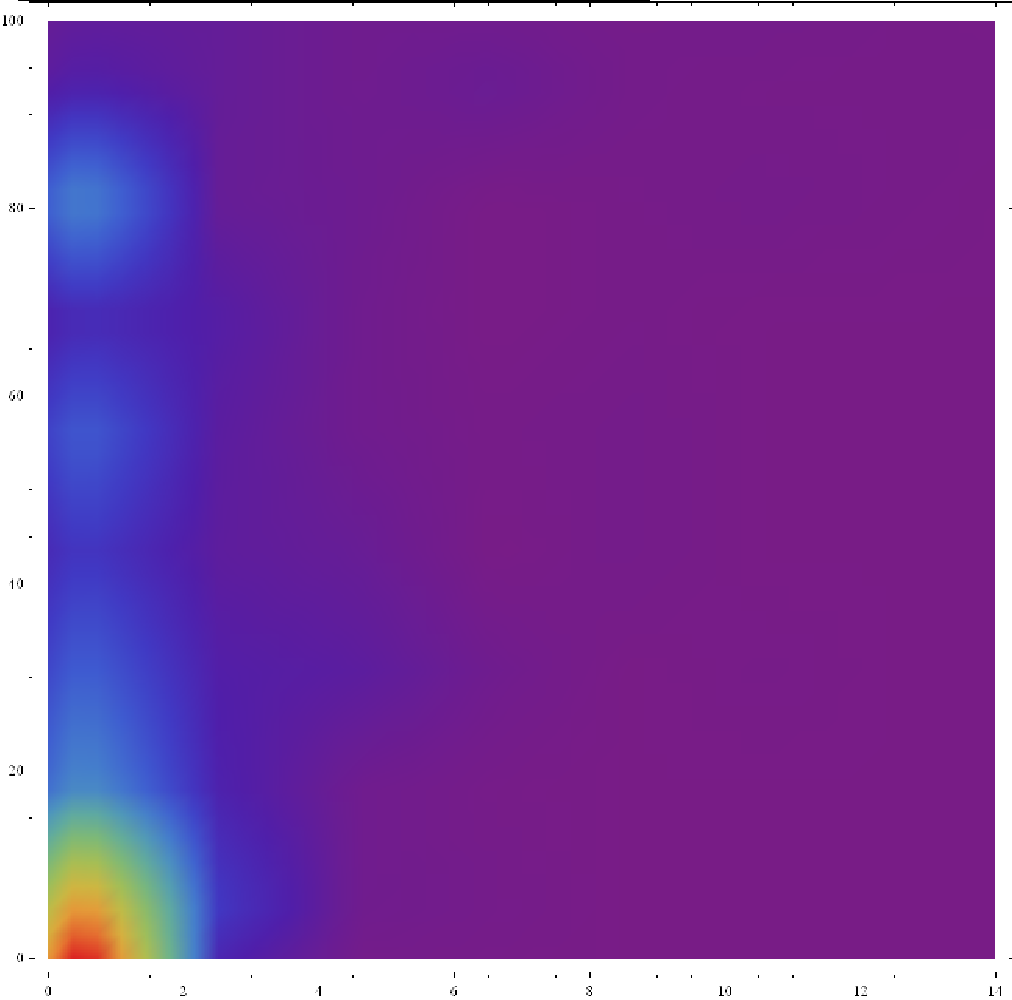}
{\textit{h. sapiens} \protect\cite{rual}}
\end{minipage}
\begin{minipage}[b]{0.3\linewidth}
\includegraphics[width = \linewidth]{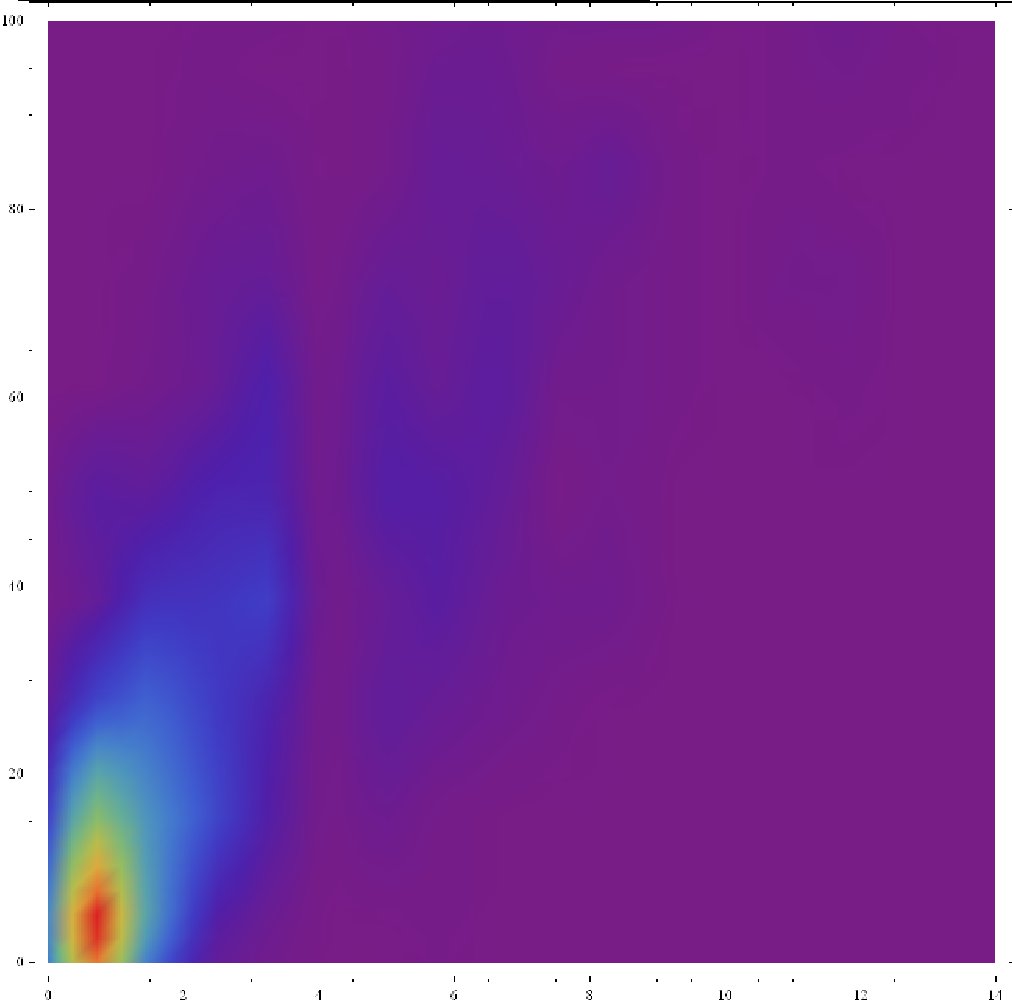}
{\textit{p. falciparum} \protect\cite{lacount}}
\end{minipage}
\begin{minipage}[b]{0.3\linewidth}
\includegraphics[width = \linewidth]{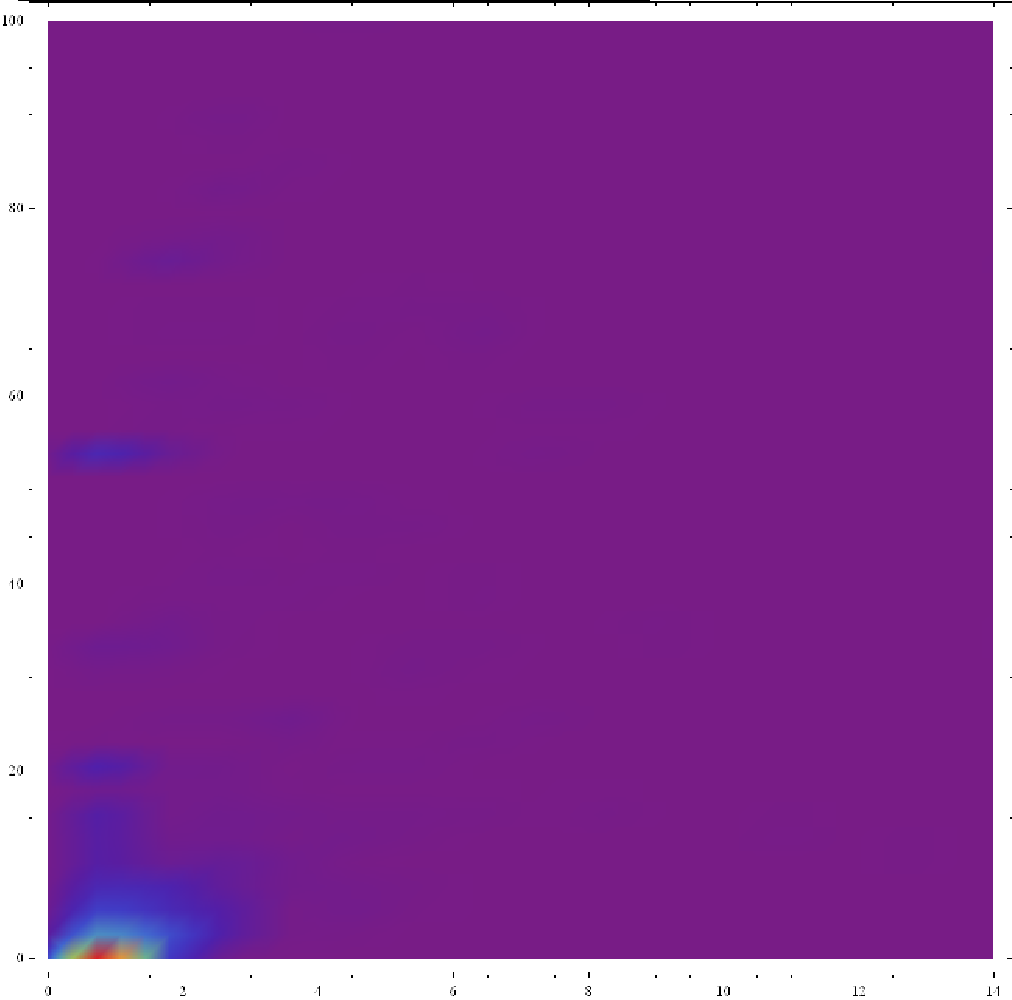} 
{\textit{s. cerevisiae} \protect\cite{ito}}
\end{minipage}\vsp
\caption{
This plot is provided in order to illustrate generalised degree distributions that appear in real biological networks. These plots are cross-referenced with the quantitative results in figure \ref{fig:complexity}.
The heatmaps are smoothed; the $x$ axis corresponds to the first degree and the $y$ axis corresponds to the second degree; the intensity of the colour indicates the observed frequency of that pair of values. }
\label{fig:biological_heatmaps}
\end{figure}

\begin{figure}[ht]
\vspace*{-6mm}
\hspace{-30pt}
\includegraphics[width = 1.1\linewidth]{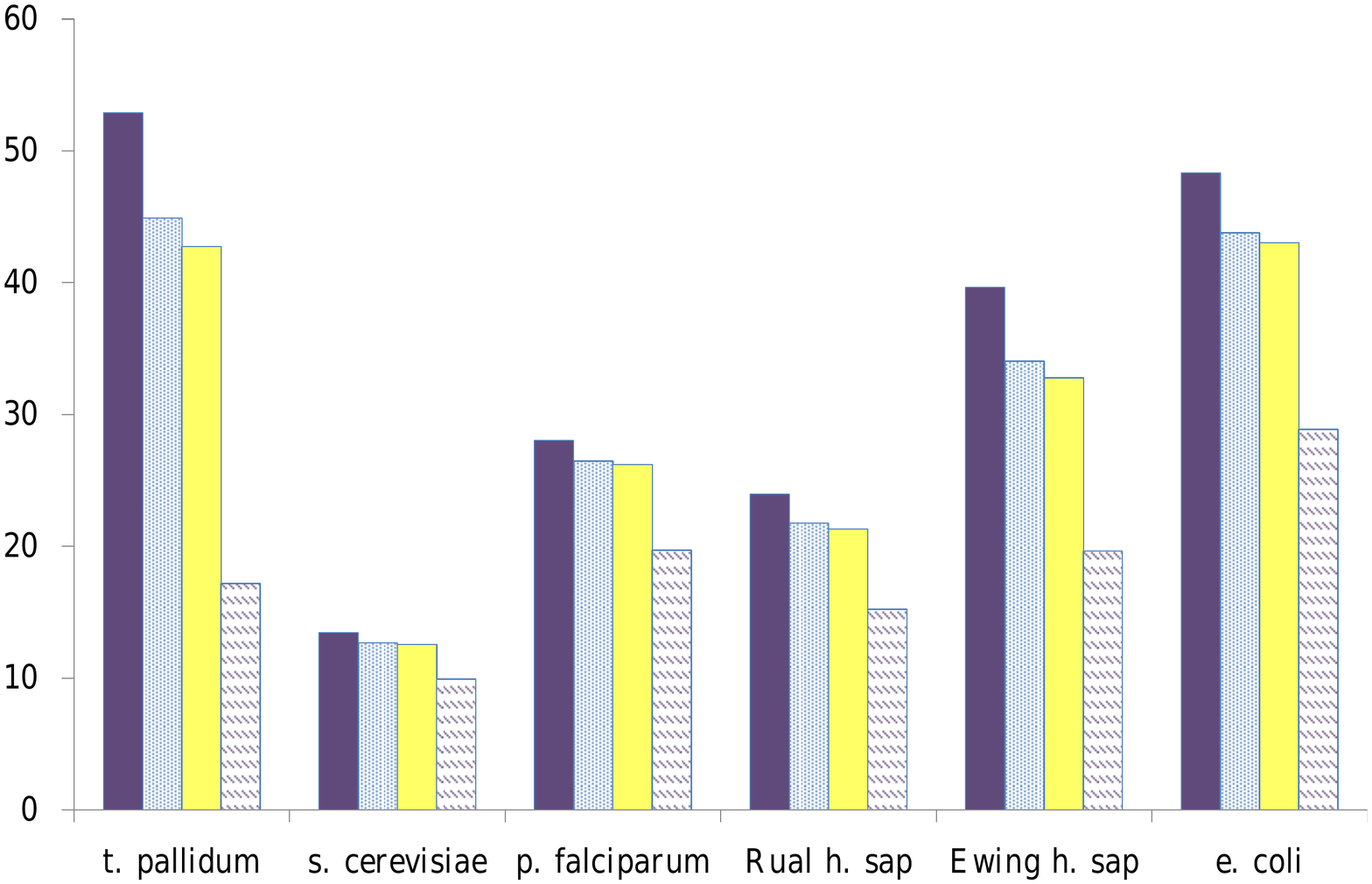}
\vspace{-35pt}
\caption{Results from applying the equations \eref{eq:entropy_final} for the entropy of random graph ensembles - where the constraints are taken to match the relevant topological observables of networks from 
\protect\cite{tre, ito, lacount, rual, ewing, arifuz}. From left to right the bars correspond to entropy per node of random graph ensembles tailored to match: average degree, degree distribution, degree-degree correlation and generalised degrees
}
\label{fig:generalised_degrees_results}
\end{figure}

Specifying the generalised degree distribution is an onerous restriction on a random graph ensemble. This was justified intuitively in section \ref{sec:GD_Intro}. As an illustration, Figure \ref{fig:generalised_degrees_results} shows this numerically by applying equation (\ref{eq:entropy_final}) to random graph ensembles tailored to the topology of real biological networks, whose generalised degrees are shown in Figure \ref{fig:biological_heatmaps}. \ref{sec:algortithm} gives some details of implementation. 

\begin{figure}[ht]
\vspace*{-5mm}
\centering
\includegraphics[width = \linewidth]{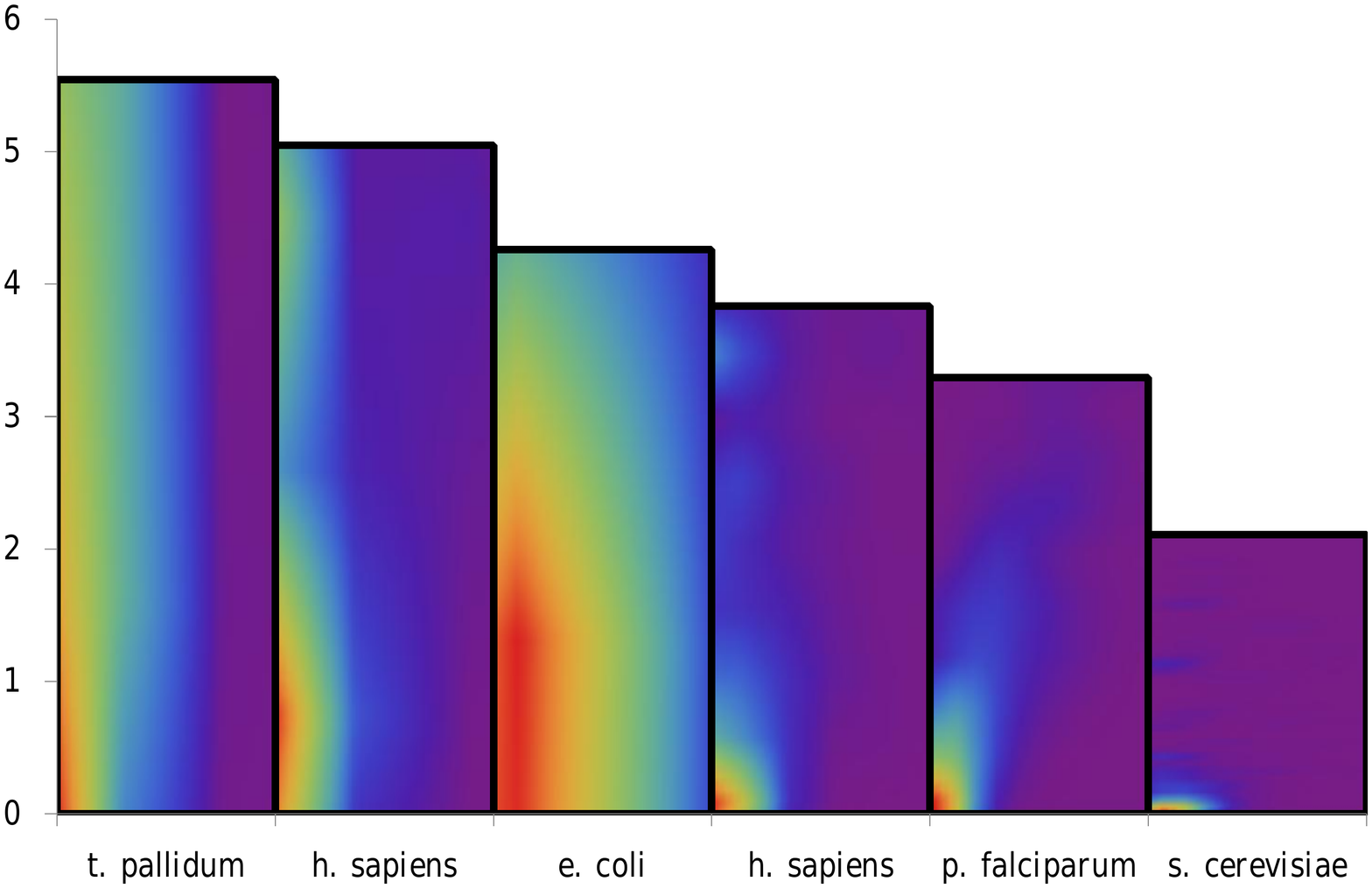}
\vspace*{-10mm}

\caption{This plot compares the calculated value for the complexity associated with the generalised degree constraint with the appearance of the generalised degree heatmap plot. From left to right along the $x$-axis the bars correspond to data from \protect\cite{tre,ewing, arifuz, rual, lacount, ito}. The $y$-axis gives the complexity per bond associated with the `generalised degrees' constraint (as taken from the corresponding dataset).  This data is equivalent to subtracting the level of the fourth bar from the level of the second bar in figure \ref{fig:generalised_degrees_results} and then normalising by dividing by the average degree $\bar{k}$.  The bars are reordered from largest to smallest for easier comparison. By inspection, it can be seen that there is a trend in the images which corresponds to intuitive expectation. The heatmaps are smoothed; the $x$ axis corresponds to the first degree and the $y$ axis corresponds to the second degree; the intensity of the colour indicates the likelihood of that pair of values.  }
\label{fig:complexity}
\end{figure}
The results show that, generally, incorporating the generalised degree constraint in the definition of the ensemble results in substantial reduction in the associated Shannon entropy. The magnitude of this reduction correlates well with a qualitative inspection of the heatmaps of the associated generalised degrees. Specifically, comparing  qualitatively confirms that there is greater complexity due to generalised degree where the generalised degree shows greater divergence from the maximum entropy simple degree distribution. Figure \ref{fig:complexity} combines Figure \ref{fig:biological_heatmaps} with Figure \ref{fig:generalised_degrees_results} to allow for direct visual comparison of the qualitative features of the generalised degree distribution obtained from these datasets with the quantitative value for the complexity associated with the same distribution (based on equation \eref{eq:entropy_final}).

\section{Conclusion}

We have derived, in leading order, an expression for the Shannon entropy of random graph ensembles constrained by the number of direct neighbours and the number of neighbours within two steps.

Approaching the calculation directly is known to lead to a self-consistency equation which resists solution (e.g. \cite{rogers2010spectral, Coolen09}). In this paper, this difficulty was avoided by taking the conceptual step of identifying that the objects appearing in the final unresolved term precisely enumerate possible allocations of neighbouring degrees. By excluding non-graphical cases, the series can be summed in terms of a topological observable: the degree-degree correlation. This provides a neat link to previous work, and shows that generalised degrees are indeed the next step in the sequence of hierarchical topological constraints studied by e.g. \cite{Coolen09}. 

The exclusion of the non-graphical cases was not able to be rigorously justified. However, it is within the spirit of the problem, and the final expression performs well under testing. The final expression takes the form of a term expressing the entropy of an ensemble constrained by a specified number of neighbours, and the specified degrees of those neighbours, plus a term reflecting the configurational entropy. The configurational entropy arises from the number of different ways that a second neighbourhood of size $m$ can be made up from the degrees of the $k$ neighbours of the node. To evaluate this factor accurately, graphicality would need to be incorporated into the analysis. Nonetheless, an upper-bound expression is provided, which already demonstrates a very material reduction in entropy under this constraint. 

In this paper, the constraint was expressed in terms of the total number of first neighbours and second neighbours. Given the way that the result appears to enumerate all possible wirings, it would be interesting to directly calculate the entropy of a random graph ensemble constrained by specifying the composition $\{ k; \xi_1... \xi_{k} \}$ of the neighbourhood of every node. This is a subset of an ensemble which has generalised degree $(k,m)$ for $m = \sum_{s=1}^k \xi_s$ specified for every node. One would expect to be able to relate one to the other, by viewing the generalised degree ensemble as the union of all the ensembles which have compliant neighbourhood compositions. 

Another natural generalisation would be to consider neighbourhoods with a wider radius: for example, specifying the number of neighbours within one, two and three steps of every node. The algebra would be quite complicated, but the substance of the result is likely to be the same: relating the number of neighbours within successively larger number of steps with the correlations in the sizes of overlapping neighbourhoods. 

These tools are relevant to the design of network models. The insight of the practitioner can define topological properties which are viewed to be critical to the design and function of the network. A random graph ensemble defined to meet these topological properties provides a pool of null-models - parallel universes where the network could plausibly function in the same way with different topology. 
Well defined null models provide a valuable check and challenge to numerical work. The Shannon entropy provides a rigorous quantitative measurement of the specificity of a model incorporating these constraints. 

The challenge in modelling a real network is to isolate the essential features, but to allow sufficient residual freedom. For any particular network, it would be possible to define a tightening sequence of topological descriptors, which will ultimately define a family of one. With the generalised degree constraint, we have taken a step closer to providing a set of descriptors that nearly uniquely define the network.  It is a stimulating alternative viewpoint to consider a network to be described by its topological constraints, rather than the more traditional connectivity list. 

\section*{Acknowledgements}

ESR gratefully acknowledges financial support from the Biotechnology and Biological Sciences Research Council of the United Kingdom. 

\section*{References}

\bibliography{gd3}{}
\bibliographystyle{vancouver}

\clearpage

\appendix

\section{Initial canonical steps to calculate Shannon entropy of random graph ensembles constrained with a specified general degree distribution}
\label{sec:calculation2}

Please refer to Section \ref{sec:definitions} for starting point of this calculation and preliminary definitions. This section outlines the first steps of the solution, which are common with e.g. \cite{Annibale09}.

\subsection{Expand the expression in order to identify the complicated term}

We
work out (\ref{eq:S}) for the ensemble (\ref{eq:ensemble}), focusing on the  leading order in $N$.
We demand that the distribution $p(k,m)$ from which the local topology characteristics $(k_i,m_i)$ are drawn does not depend on $N$, so
that with the law of large numbers one obtains
\begin{eqnarray}
S&=&N^{-1}\sum_{\bc}p(\bc)\log Z(\bk(\bc),\bm(\bc))
-N^{-1}\sum_i \sum_{\bc}p(\bc)\log p(k_i(\bc),m_i(\bc))
\nonumber
\\
&=&
\sum_{\bk,\bm} p(\bk,\bm) N^{-1}\log Z(\bk,\bm)
-\sum_{\bk,\bm}  p(\bk,\bm) \Big(N^{-1}\sum_i \log p(k_i,m_i)\Big)
\nonumber
\\
&=&
\sum_{\bk,\bm} p(\bk,\bm) N^{-1}\log Z(\bk,\bm)
- \sum_{k,m}p(k,m) \log p(k,m)+\epsilon_N
\end{eqnarray}
where $\lim_{N\to\infty}\epsilon_N=0$.
\vsp

We next focus on the first term. We abbreviate $\overline{k}=N^{-1}\sum_i k_i$ and define a Poissonnian measure
\begin{eqnarray}
w(\bc|\overline{k})&=& \prod_{i<j}\Big[
\frac{
		\overline{k}
	}
	{
		N
	}
\delta_{c_{ij},1}+
(1-
\frac{
		\overline{k}
	}
	{
		N
	}
)\delta_{c_{ij},0}\Big]
\end{eqnarray}
This measure is normalized, and depends on $\bc$ only via its average degree $\overline{k}(\bc)$, since
\begin{eqnarray}
w(\bc|\overline{k})&=& \Big[
\frac{
		\overline{k}
	}
	{
		N		
	}
\Big]^{\sum_{i<j}c_{ij}}
\Big[1-
\frac{
		\overline{k}
	}
	{
		N
	}
\Big]^{\sum_{i<j}(1-c_{ij})}
\nonumber
\\
&=&\Big[1-
\frac{
		\overline{k}
	}
	{
		N
	}
\Big]^{
		\frac{1}{2}
		N(N-1)
		}
\Big[
\frac{
		\overline{k}/N
	}
	{1 \!-\!\overline{k}/N
	}
\Big]^{
		\frac{1}{2}
	\sum_{ij}c_{ij}}
\nonumber
\\
&=&\Big[1-\frac{\overline{k}}{N}\Big]^{\frac{1}{2}N(N-1)}
\Big[\frac{\overline{k}/N}{1\!-\!\overline{k}/N}\Big]^{\frac{1}{2}N\overline{k}(\bc)}
\end{eqnarray}
Let us write the latter expression as $W(\overline{k},\overline{k}(\bc))$.
We may therefore write $Z(\bk,\bm)$ as
\begin{eqnarray}
Z(\bk,\bm)&=&\sum_{\bc}
\frac{
		w(\bc|\overline{k})
	}
	{
		W(\overline{k},\overline{k}(\bc))
	}
\prod_i\Big(\delta_{k_i,k_i(\bc)}\delta_{m_i,m_i(\bc)}\Big)
\nonumber
\\
&=&
\frac{
		1
	}
	{
		W(\overline{k},\overline{k})
	}
\sum_{\bc}w(\bc|\overline{k})\prod_i\Big(\delta_{k_i,k_i(\bc)}\delta_{m_i,m_i(\bc)}\Big)
\end{eqnarray}
We thus obtain for the entropy the following expression, with $\bra \ldots\ket_\kappa$ denoting an average over the Poissonian measure with average connectivity $\kappa$, i.e. $\bra f(\bc)\ket_\kappa=\sum_{\bc}w(\bc|\kappa)f(\bc)$,
\begin{eqnarray}
S&=&
\frac{1}{N}
\sum_{\bk,\bm} p(\bk,\bm)\log \Big\bra \prod_i\Big(\delta_{k_i,k_i(\bc)}\delta_{m_i,m_i(\bc)}\Big)\Big\ket_{\overline{k}}
\nonumber\\
&&
-\frac{1}{N}
\sum_{\bk,\bm} p(\bk,\bm) \log W(\overline{k},\overline{k})
- \sum_{k,m}p(k,m) \log p(k,m)+\epsilon_N
\nonumber
\\
&=&
\frac{1}{2}
\bra k\ket[1+ \log (N/\bra k\ket)]
- \sum_{k,m}p(k,m) \log p(k,m)
\nonumber
\\
&&+\frac{1}{N}
\sum_{\bk,\bm} p(\bk,\bm)\log \Big\bra \prod_i\Big(\delta_{k_i,k_i(\bc)}\delta_{m_i,m_i(\bc)}\Big)\Big\ket_{\overline{k}}
+\epsilon_N
\end{eqnarray}

\subsection{Expressing Kronecker $\delta$ in Fourier Form}
All the complexities of the problem are contained in the last line, which we abbreviate as
\begin{eqnarray}
\phi(\bk,\bm)&=&\frac{1}{N}
\sum_{\bk,\bm} p(\bk,\bm)\log \Big\bra \prod_i\Big(\delta_{k_i,k_i(\bc)}\delta_{m_i,m_i(\bc)}\Big)\Big\ket_{\overline{k}}
\end{eqnarray}
Using integral representations of the Kronecker $\delta$-functions allows us to write this nontrivial contribution to the entropy as
\begin{eqnarray}
\phi(\bk,\bm)
&=&\frac{1}{N}
\sum_{\bk,\bm} p(\bk,\bm)\log
\int_{-\pi }^{\pi }\!\prod_{i}\Big[
\frac{
		\rmd \omega_i\rmd\psi_i
	}
	{
		4\pi^2
	}
\rme^{\rmi(\omega_i k_i+\psi_im_i)}\Big]
 L(\bomega,\bpsi)
\\
L(\bomega,\bpsi)&=&
\sum_c
	\prod_{i<j}
		\left[
			\frac{\bar{k}}{N}
			\delta_{c_{ij},1}		
		+	(1 - \frac{\bar{k}}{N})\delta_{c_{ij},0}	
		\right]
	\rme ^ {-\rmi \sum_{ij}c_{ij} (\omega_i + \phi_i k_j)}
\end{eqnarray}
with $\bomega=(\omega_1,\ldots,\omega_N)$ and $\bpsi=(\psi_1,\ldots,\psi_N)$.

By symmetry it is clear that $\sum_{i \neq j}c_{ij}(\omega_i + \phi_i k_j) = \sum_{i<j}c_{ij}(\omega_i + \phi_i k_j + \omega_j + \phi_j k_i) $ so

\begin{eqnarray*}
L(\bomega,\bpsi)&=&
\sum_c
	\prod_{i<j}
		\left[
			\frac{\bar{k}}{N}
			\delta_{c_{ij},1}		
		+	(1 - \frac{\bar{k}}{N})\delta_{c_{ij},0}	
		\right]
	\exp [-\rmi c_{ij} (\omega_i + \phi_i k_j + \omega_j + \phi_j k_i)]
\end{eqnarray*}

using the action of the $\delta$s, we reach:

\begin{eqnarray}
L(\bomega,\bpsi)&=&
\sum_c
	\prod_{i<j}
		\left[
			\frac{\bar{k}}{N}
			[
				\delta_{c_{ij},1}	\rme ^ {-\rmi (\omega_i + \phi_i k_j + \omega_j + \phi_j k_i)} - \delta_{c_{ij},0}
			]	
		+	\delta_{c_{ij},0}	
		\right]
\end{eqnarray}

It is valid to exchange to exchange the order of the summation and the product, since we are in an 'all possible combinations' combinatorial shape. 

\begin{eqnarray}
L(\bomega,\bpsi)
&=&
	\prod_{i<j}
	\sum_{c_{ij}}
		\left[
			\frac{\bar{k}}{N}
			[
				\delta_{c_{ij},1}	\rme ^ {-\rmi (\omega_i + \phi_i k_j + \omega_j + \phi_j k_i)} - \delta_{c_{ij},0}
			]	
		+	\delta_{c_{ij},0}	
		\right]
\\ \nonumber
&=&
	\prod_{i<j}
		\left[
			\frac{\bar{k}}{N}
			[
				\rme^{-\rmi (\omega_i + \phi_i k_j + \omega_j + \phi_j k_i)} - 1
			]	
		+	1	
		\right]
\end{eqnarray}

Now express $L(\bomega,\bpsi)$ as the exponential of the logarithm. Taking the product outside the logarithm as a sum, and taking the leading order in $N$ of the Taylor expansion for each term, it immediately follows that 

\begin{eqnarray}
L(\bomega,\bpsi)
&=&
	\exp 
	\left[
		\sum_{i \neq j}
			\frac{\bar{k}}{2N}
			[
				\rme^{-\rmi (\omega_i + \phi_i k_j + \omega_j + \phi_j k_i)} - 1
			]	
		+
		\order(N^{-2})
	\right]
\end{eqnarray}

\subsection{Constructing special functions to bring down the site-specific variables}

We now have a sum over site-specific variables. 
Define 
\begin{eqnarray}
P(\bomega,\bpsi, \bk | \vec{\omega}, \vec{\phi}, \vec{k}) \equiv \frac{1}{N} \sum_i \delta(\omega - \omega_i) \delta(\phi - \phi_i) \delta_{k, k_i}
\end{eqnarray}
%
By construction $\sum_k \int P(\omega, \phi, k | \vec{\omega}, \vec{\phi}, \vec{k}) \rmd \omega \rmd \phi \equiv 1$. So we can write
\begin{eqnarray*}
\hspace*{-60pt}
L(\bomega,\bpsi)
&=
\exp
\left[
  \int  \rmd \vec{\omega} \rmd \vec{\omega^\prime} \sum_{k, k^\prime} 
  		P(\omega, \phi, k | \vec{\omega}, \vec{\phi}, \vec{k}) 
  	    P(\omega^\prime, \phi^\prime, k^\prime | \vec{\omega}, \vec{\phi}, \vec{k}) 
  	    \left[ \frac{\bar{kN}}{2} \rme^{-\rmi (\omega + \phi k^\prime + \omega^\prime + \phi^\prime k)}	\right]
  	    \rme^{-\frac{\bar{k}N}{2}}
\right]
\end{eqnarray*}
Before we proceed further we need to discretise $L$ - this makes it easier to see how the next step works. 
\begin{eqnarray*}
\hspace*{-60pt}
L(\vec{\omega})=
\exp
\left[
 \sum \Delta \omega \Delta \omega^\prime \Delta \phi \Delta \phi^\prime
P(\omega, \phi, k | \vec{\omega}, \vec{\phi}, \vec{k}) 
P(\omega^\prime, \phi^\prime, k^\prime | \vec{\omega}, \vec{\phi}, \vec{k}) 
\left[ \frac{\bar{kN}}{2} 
\exp\left[-\rmi (\omega + \phi k^\prime + \omega^\prime + \phi^\prime k)	\right]
  	    \rme^{-
  	    		\frac{\bar{k}N}{2}
  	    	}
\right]
\right]
\end{eqnarray*}
Now - for every value of $\omega, \phi, k$ apply the following relation
$\int \rmd P(\omega, \phi, k) \, \delta \left[P(\omega, \phi, k) - P(\omega, \phi, k | \vec{\omega}, \vec{\phi}, \vec{k})  \right] $ in order to bring our the dependence on $\bc$. This lets us replace every instance of $P(\omega, \phi, k | \vec{\omega}, \vec{\phi}, \vec{k})$ with $P(\omega, \phi, k)$. 
\begin{eqnarray*}
L(\vec{\omega})=\prod_{\omega, \omega^\prime, \phi, \phi^\prime, k, k^\prime}
\int
\rmd P(\omega, \phi, k) \, \delta \left[P(\omega, \phi, k) - P(\omega, \phi, k | \vec{\omega}, \vec{\phi}, \vec{k})  \right] \\
\exp \left\lbrace \Delta \omega \Delta \omega^\prime \Delta \phi \Delta \phi^\prime
P(\omega, \phi, k ) 
P(\omega^\prime, \phi^\prime, k^\prime) 
\left[ \frac{\bar{kN}}{2} \rme^{-\rmi (\omega + \phi k^\prime + \omega^\prime + \phi^\prime k)}	\right]
  	    \rme^{-\frac{\bar{k}N}{2}}
\right\rbrace
\end{eqnarray*}
Now we are going to express $\delta\left[ P(\omega, \phi, k) - P(\omega, \phi, k | \vec{\omega}, \vec{\phi}, \vec{k})  \right]$ in Fourier form $$\int \rmd x 
\,	
	\frac{
			\exp\left[{
					\rmi x \left(P(\omega, \phi, k) - P(\omega, \phi, k | \vec{\omega}, \vec{\phi}, 								\vec{k})  \right)
					}\right]
		}
		{2 \pi}$$
		 - but in place of $x$ we are going to use $x= N \Delta \omega \Delta \phi \hat{P}(\omega, \phi, k)$ which also means we can substitute $\frac{\rmd x}{\rmd \hat{P}} = N \Delta \omega \Delta \phi$.

At this point we have the opportunity to spot that we have achieved a standard path integral form where we can write $\prod_{\omega, \phi, k} \frac{N \Delta \phi \Delta \omega}{2 \pi} \rmd {P}(\omega, \phi, k)\rmd \hat{P}(\omega, \phi, k) $ as $\{ \rmd {P} \rmd \hat{P}\}$. 

We have arrived at the following form
\begin{eqnarray}
L(\vec{\omega}) = \int 
\left\lbrace \rmd {P} \rmd \hat{P} \right\rbrace
\rme^{
		\rmi N 
		\int_{- \pi}^{\pi} 
			\rmd {\omega}  \rmd {\phi} 
			\sum_k 
				\hat{P}(\omega, \phi, k) 
				\left( 
					{P}(\omega, \phi, k)
					-{P}(\omega, \phi, k | \vec{\omega}, \vec{\phi}, k )\right)
		}
\times
\rme^
{
	\frac{-\bar{k} N}{2}
} \\ 
\exp
{
\left[
	\frac{-\bar{k}N}{2}
	\int \rmd {\omega}  \rmd {\phi} \rmd {\omega^\prime}  \rmd {\phi^\prime} 
	\sum_{k, k^\prime}
		 {P}(\omega, \phi, k)  {P}(\omega^\prime, \phi^\prime, k^\prime)
		\rme^
		{
			-\rmi (\omega + \omega^\prime + \phi k^\prime + \phi^\prime k )		
		}
\right]
}
\end{eqnarray}
and restore the original delta function meaning of ${P}(\omega, \phi, k | 						\vec{\omega}, \vec{\phi}, k )$. 

\subsection{Law of large numbers}

Recall that our form of interest was
\begin{eqnarray}
\phi = N^{-1} \sum_{k, m} p (k, m) 
\log
\left[ 
			\int \prod_i 
				\frac
				{
					\rmd \bomega_i  
						\rme^{
						\rmi 
								\bomega_i \cdot {\bk_i} 
							}
							L(\bomega)							
				} 				
				{4 \pi^2}
\right] 
\end{eqnarray}
We can pull out the site-specific variable 

\begin{eqnarray}
\int \prod_i \left[ \frac
					{\rmd \bomega_i}
					{4 \pi^2}
				 \rme^{\rmi(\vec{\omega_i}\cdot \vec{k_i} - \hat{P}(\vec{\omega_i}, \vec{k_i}))} \right] 
\end{eqnarray}
By symmetry this expression should not change under exchange of nodes, so we can re-write it as
\begin{eqnarray}
\prod_i \int \left[ \frac{\rmd \bomega}{4 \pi^2} \rme^{\rmi(\bomega.k_i - \hat{P}(\bomega, k_i))} \right] = 
\exp 
	\left(
			\sum_i 
				\log 
						\int
						\frac{\rmd \bomega}
								{4 \pi^2} \rme^{\rmi \left(\bomega.k_i - \hat{P}(\bomega, k_i) \right)} 
	\right)
\end{eqnarray}
which by the law of large numbers can be expressed as
\begin{eqnarray}
\exp \left[N \sum_{\bk} 
 p(\bk) \log \int 
\frac{\rmd \bomega}{4 \pi^2}
\rme^{ \rmi ( \bomega \cdot \bk - \hat{P} ) } \right]
\end{eqnarray}
\subsection{Saddle point form}

Which leads us to the following form for $\phi$
\begin{eqnarray}
\phi = \lim_{N \rightarrow \infty} N^{-1}
 \sum_{\bk} p(\bk) \log  \int \left\lbrace\rmd P \rmd \hat{P} \right\rbrace 
\rme^{N \Psi [P, \hat{P}]}
\end{eqnarray}
where 
\begin{eqnarray}
\label{eq:gd_big_psi}
\Psi [P, \hat{P}]
 &= &\rmi \int^\pi_{-\pi} \rmd \bomega \sum_{\bk} P(\bomega, \bk) \hat{P}(\bomega, \bk ) \\ \nonumber
&&+ \frac{\bar{k}}{2}
	\int \rmd \bomega \rmd \bomega^\prime \sum_{\bk, \bk^\prime} P(\bomega, \bk)P(\bomega^\prime, \bk^\prime)\, \exp[-\rmi (\omega + \omega^\prime + \phi k^\prime_1+ \phi^\prime k_1)] \\ \nonumber
&&- \frac{\bar{k}}{2}
+ \sum_{\bk} p(\bk) \log \int^\pi_{-\pi} \frac{\rmd \bomega}{4 \pi^2} \,\rme^{ \rmi (\bomega \cdot  \bk - \hat{P}(\bomega, \bk) )}
\end{eqnarray}

Apart from the exponential in the second term, it is important to observe that all the elements of this expression are two component vectors: $\bk=(k,m)$ represents the first and second neighbourhood of a node, $\bomega = (\omega, \phi)$ is the two component integration variable. 

By a standard saddle point argument, the solution can be found by
\begin{eqnarray}
\phi = \extr_{\{ P, \hat{P}\}} \Psi \left[ P, \hat{P}\right]
\end{eqnarray}
via functional differentiation

\subsection{Finding the extremum with functional differentiation}

First make a substitution for easier manipulation $Q = \rme^{-\rmi \hat{P}}$ so that we can write
\begin{eqnarray}
\Psi [P, \hat{P}]
 &= & \int^\pi_{-\pi} \rmd \bomega \sum_{\bk} P(\bomega, \bk)\log{Q}(\bomega, \bk ) \\ \nonumber
&&+ \frac{\bar{k}}{2}
	\int \rmd \bomega \rmd \bomega^\prime \sum_{\bk, \bk^\prime} P(\bomega, \bk)P(\bomega^\prime, \bk^\prime) \rme^{-\rmi (\omega + \omega^\prime + \phi k^\prime+ \phi^\prime k)}
- \frac{\bar{k}}{2} \\ \nonumber
&&+ \sum_{(k,m)} p(k,m) \log \int^\pi_{-\pi} \frac{\rmd \bomega}{4 \pi^2} \rme^{ \rmi \omega . \bk } {Q}(\bomega, \bk )
\end{eqnarray}

Perform functional differentiation and setting $\frac{\delta \Psi}{\delta P} = \frac{\delta \Psi}{\delta Q} = 0$
\begin{eqnarray}
\log{Q}(\bomega, \bk ) &=&  \bar{k} \int \rmd \bomega^\prime \sum_{k^\prime} P(\bomega^\prime, k^\prime)
\rme^{- \rmi(\omega + \omega^\prime + \phi k^\prime + \phi^\prime k )} \\ \nonumber
P(\bomega, \bk ) &=& Q (\bomega, \bk )
\frac
{
	p(\bk) \rme^{\rmi \bomega \cdot  \bk}
}
{
	\int \rmd \bomega Q(\bomega, \bk )\rme^{\rmi \bomega \cdot \bk}
}
\end{eqnarray}
Eliminate $P(\bomega, \bk ) $ and define $\gamma$ to satisfy 
\begin{eqnarray}
Q &=& \exp 
\left( 
     \bar{k} \sum_{\zeta \in \mathbb{N}^2} \gamma(\bk, \zeta) \rme^{-\rmi \bomega \cdot \zeta}
\right)
\end{eqnarray}
which means that we require
\begin{eqnarray*}
\sum_{\bmu \in \mathbb{N}^2} \gamma(\bk, \bmu) \rme^{-\rmi \bomega \cdot \bmu}
&=&
\sum_{k^\prime}
\frac
{
	\int \rmd \bomega^\prime  
	\exp\left(
				\bar{k}\sum_{\zeta \in \mathbb{N}^2} \gamma(\bk^\prime, \zeta) \rme^{-\rmi \bomega^\prime \cdot \zeta}	
			\right)
			\rme^{\rmi \bomega^\prime \cdot \bk^\prime}
	\rme^{- \rmi(\omega + \omega^\prime + \phi k^\prime + \phi^\prime k )} 
	p(\bk^\prime)
}
{
	\int \rmd \bomega^{\prime \prime} 
	\exp\left(
				\bar{k}\sum_{\zeta \in \mathbb{N}^2} \gamma(\bk^\prime, \zeta) \rme^{-\rmi \bomega^{\prime \prime} \cdot \zeta}	
			\right)
			\rme^{\rmi \bomega^{\prime \prime} \cdot \bk^\prime}
}\\ \nonumber
\end{eqnarray*}
Compare coefficients of the free $\omega$ to reach
\begin{eqnarray}
\label{eq:gd_gamma1}
\hspace*{-60pt}
\gamma(\bk, \bmu)
=
\delta_{\mu_1,1}
\sum_{k^\prime} 
\frac
{
	\int \rmd \bomega^\prime 
	\exp\left(
				\bar{k}\sum_{\bzeta \in \mathbb{N}^2} \gamma(\bk^\prime, \bzeta) \rme^{-\rmi \bomega^\prime \cdot \bzeta}	
			\right)
			\rme^{\rmi \bomega^\prime \cdot \bk^\prime}
	\rme^{- \rmi(\omega^\prime + \phi^\prime k )} 
	p(\bk^\prime) \delta_{\mu_2,k_1^\prime}
}
{
	\int \rmd \bomega^{\prime \prime} 
	\exp\left(
				\bar{k}\sum_{\bzeta \in \mathbb{N}^2} \gamma(\bk^\prime, \bzeta) \rme^{-\rmi \bomega^{\prime \prime} \cdot \bzeta}	
			\right)
			\rme^{\rmi \bomega^{\prime \prime} \cdot \bk^\prime}
}\\ \nonumber
\end{eqnarray}
Do the integral on the bottom line by doing a series expansion of the exponential
\begin{eqnarray*}
\hspace*{-60pt}
\int
\rmd \bomega \,
\rme^{\rmi \bomega \cdot \bk^\prime}
\exp 
\left( 
\bar{k} 
\sum_{\bzeta} \gamma(\bk^\prime, \bzeta)
\rme^{-\rmi \bomega \cdot \bzeta}
\right)
&=
\sum_{r=1}^\infty \frac{\bar{k}}{r!}
\int
\rmd \bomega
\rme^{\rmi \bomega \cdot \bk^\prime}
\left( 
\bar{k} 
\sum_{\bzeta} \gamma(\bk^\prime, \bzeta)
\rme^{-\rmi \bomega \cdot \bzeta}
\right)^r \\
&=
\sum_{r=1}^\infty \frac{\bar{k}^r}{r!}
4 \pi^2
\sum_{\bzeta^1....\bzeta^r \in \mathbb{N}^2} 
\prod_{s=1}^r 
\gamma(\bk, \bzeta^s)
\delta_{k_1^\prime, \sum_s \zeta_1^s}
\delta_{k_2^\prime, \sum_s \zeta_2^s}
\end{eqnarray*}
but this can be simplified further if we look back to equation \eref{eq:gd_gamma1}, which requires $\bzeta_1 = 1$ for $\gamma$ to be non zero. Hence the bottom line of equation \eref{eq:gd_gamma1} is found to be 
\begin{eqnarray*}
\mbox{Bottom line}
&=&
\frac{
		4 \pi^2 \bar{k}^{k_1^\prime}
	}
	{
		k_1^\prime!
	}
\sum_{\bzeta^1....\bzeta^k_1 \in \mathbb{N}^2} 
\prod_{s=1}^{k_1^\prime} 
\gamma(\bk, \bzeta^s)
\delta_{k_2^\prime, \sum_s \bzeta_2^s}
\end{eqnarray*}
and by the same process the top line of equation \eref{eq:gd_gamma1} is found to be 
\begin{eqnarray*}
\mbox{Top line}
&=&
\frac{
		4 \pi^2 \bar{k}^{k_1^\prime-1}
	}
	{
		(k_1-1)!
	}
\sum_{\bzeta^1....\bzeta^{k_1^\prime-1} \in \mathbb{N}^2} 
\prod_{s=1}^{k_1^\prime-1} 
\gamma(\bk, \bzeta^s)
\delta_{k_2^\prime - k_1, \sum_s \zeta_2^s}
\end{eqnarray*}

Now we have an expression for $\gamma$ at the point where $P$ and $\hat{P}$ maximise $\Psi$. Write
$$ \Gamma = \exp \left(\bar{k} \sum_\xi \gamma(k,\xi) \rme^{- \rmi \omega \xi} \right)$$
so that
$$P(\omega, \bk) = 
\frac{
		p(\vec{k} ) \rme^{\rmi \omega k} \Gamma(\omega, k))
	}
	{
		\int_{-\pi}^{\pi} \rmd \bomega \ \rme^{\rmi \omega k} \Gamma(\omega, k))
	} $$
from which it follows that
$$ 
\Psi = 
- \frac{
\int \rmd \omega  \sum_{\vec{k}} p(\vec{k})\rme^{\rmi \omega k} \Gamma(\omega, k) \log \Gamma(\omega, k)  
}
{
2 \int \rmd \omega^\prime  \sum_{\vec{k}} p(\vec{k})\rme^{\rmi \omega^\prime k} \Gamma(\omega^\prime, k) 
}
+ \sum_{\vec{k}} p(\vec{k}) \log \int \frac{\rmd \omega}{4 \pi^2}
\rme^{\rmi \omega k} \Gamma(\omega, k)
- \frac{\bar{k}}{2}
$$
It is now necessary to evaluate all of the components of this expression
\begin{eqnarray*}
\int \rmd \omega
\
\rme^{\rmi \omega k} \Gamma(\omega, k)
&=& \sum_{r=0}^{\infty} \frac{\bar{k}^r}{r!}
\int \rmd \omega
\
\rme^{\rmi \omega k}
\left( \sum_{\xi} \gamma(k, \xi) \rme^{-\rmi \omega k}\right)^r \\
&=&
4 \pi^2 \left( \delta_{k,0} + \frac{\bar{k}^k}{k!} \theta(k-\frac{1}{2})\sum_{\xi_1...\xi_k}\prod_{\xi_1...\xi_k} \gamma(k,\xi_i) \delta_{m, \sum_i^k \xi_i}\right)
\end{eqnarray*}
and with the same technique
\begin{eqnarray*}
\int \rmd \omega
\
\rme^{\rmi \omega k} 
\Gamma(\omega, k)
 \log \Gamma(\omega, k)
&=& 
4 \pi^2 
\frac{\bar{k}^k}{(k-1)!}
 \theta(k-\frac{1}{2})
 \sum_{\xi_1...\xi_k}\prod_{\xi_1...\xi_k} \gamma(k,\xi_i) 
 \delta_{m, \sum_i^k \xi_i}
\end{eqnarray*}

where the $\theta$ is a heaviside function to distinguish the $k=0$ case - which will be disregarded from now on, since we have defined our ensemble to have no isolated nodes.
The first term of $\Psi$ cancels down to $\frac{- \bar{k}}{2}$. Combine this with the trailing $\frac{- \bar{k}}{2}$ and re-write as $\sum_{\vec{k}}p(\vec{k}) \log \rme^{-\bar{k}}$. Recall that $\pi(k)= \frac{\bar{k}^k \rme^{-\bar{k}}}{k!}$, and hence write the final form of $\Psi$ as 
$$
\Psi = -\sum_k p(k) \log \pi(k) + \sum_{\vec{k}} p(\vec{k}) \log \left[ 
\sum_{\xi_1...\xi_k} \prod_{\xi_1...\xi_k}
\gamma(k, \xi_i) \delta_{m, \sum \xi_i} 
\right]
$$

\subsection{Intermediate Answer}
\begin{eqnarray}
S &=& \frac{\bar{k}}{2}(1 + \log (\frac{N}{\bar{k}})) - \sum_k p(k,m) \log \frac{p(k,m)}{\pi(k)}  
\\ \nonumber &&
\sum_{(k,m)} p(k,m) \log [ \sum_{(\xi_1, ..., \xi_k)} \prod_{s=1}^k \gamma(k, \xi_s) \delta_{m, \sum^k_{i=1} \xi_i} ]
+ e_N
\end{eqnarray}
where $\gamma$ satisfies the self-consistency relation ($m$ is used instead of $k_2$ for clarity)
\begin{eqnarray}
\gamma{(\vec{k}, \vec{\xi})} 
= \delta_{\xi_1,1} 
\sum_{k^{\prime}}{\delta_{\xi_2,k^\prime_1} \frac{k^\prime_1}{\bar{k}}p(k^\prime_1) 
\left[
	\frac
	    {
	    	\sum_{\xi^1...\xi^{k^\prime-1}} 
	    		\delta_{m^\prime-k, \sum_{s=1}^{k^\prime-1}\xi_2^s}
	    		\prod_{s=1}^{k^\prime-1} \gamma\left( k^\prime, \xi^s \right)
	    }
	    {
	    	\sum_{\xi^1...\xi^{k^\prime}} 
	    		\delta_{m^\prime, \sum_{s=1}^{k^\prime}\xi_2^s}
	    		\prod_{s=1}^{k^\prime} \gamma\left( k^\prime, \xi^s \right)
	    }
\right]
}
\end{eqnarray}
This can simplify a little bit. We can observe that in fact, $\gamma$ depends on only two arguments: $k_1, \xi_2$. For convenience, relabel these as $k, k^\prime$ and rewrite the self consistency relation as
\begin{eqnarray}
\gamma({k, k^\prime}) 
= 
\sum_{{m}^{\prime}}{ \frac{k^\prime}{\bar{k}}p({k}^\prime, {m}^\prime) 
\left[
	\frac
	    {
	    	\sum_{\xi^1...\xi^{k^\prime-1}} 
	    		\delta_{m^\prime-k, \sum_{s=1}^{k^\prime-1}\xi_2^s}
	    		\prod_{s=1}^{k^\prime-1} \gamma\left( k^\prime, \xi^s \right)
	    }
	    {
	    	\sum_{\xi^1...\xi^{k^\prime}} 
	    		\delta_{m^\prime, \sum_{s=1}^{k^\prime}\xi_2^s}
	    		\prod_{s=1}^{k^\prime} \gamma\left( k^\prime, \xi^s \right)
	    }
\right]
}
\end{eqnarray}

\section{Implied generalised degree-degree correlation}
\label{sec:W_appendix}

Define
\begin{eqnarray}
W(\vec{k}, \vec{k^\prime}|c) &=& \frac{\sum_{ij} c_{ij} \delta_{\vec{k}, \vec{k}_i} 
\delta_{\vec{k}^\prime, \vec{k^\prime}_i}
}
{
\sum_{ij} c_{ij}
}
\end{eqnarray}
where $\vec{k} = (k,m)$, the generalised degree. What is the average of this quantity in our ensemble with a specified generalised degree distribution? We can calculate this by looking at
\begin{eqnarray}
W(\vec{k}, \vec{k^\prime}) &=& \lim_{N \rightarrow \infty} \sum_{\bc} prob(\bc|p(\vec{k}))W(\vec{k}, \vec{k^\prime}|c)
\end{eqnarray}
where $prob(\bc|p(\vec{k})) = \sum_{\{ \vec{k}_1...\vec{k}_N \}} 
\frac{ \prod \delta_{\vec{k}_i, \vec{k}_i(\bc)} }
{\sum_{\bc^\prime} \prod \delta_{\vec{k}_i, \vec{k}_i(\bc^\prime)} }
\prod p(\vec{k}_i)$
Unpack this expression
\begin{eqnarray*}
W(\vec{k}, \vec{k^\prime})
&=&
\lim_{N \rightarrow \infty} 
\sum_{\bc} 
\sum_{\{ \vec{k}_1...\vec{k}_N \}}
\frac{ \prod \delta_{\vec{k}_i, \vec{k}_i(\bc)} }
{\sum_{\bc^\prime} \prod \delta_{\vec{k}_i, \vec{k}_i(\bc^\prime)} }
\prod p(\vec{k}_i)
\frac{\sum_{rs} c_{rs} \delta_{\vec{k}, \vec{k}_r} 
\delta_{\vec{k}^\prime, \vec{k^\prime}_s}
}
{
\sum_{ij} c_{ij}
}\\
&=&
\frac{1}{\bar{k}N}
\sum_{rs} 
\sum_{\{ \vec{k}_1...\vec{k}_N \}}
\prod p(\vec{k}_i)
\delta_{\vec{k}, \vec{k}_r} 
\delta_{\vec{k}^\prime, \vec{k}_s} 
\frac{ \sum_{\bc} c_{rs}
\prod \delta_{\vec{k}_i, \vec{k}_i(\bc)} }
{\sum_{\bc^\prime} \prod \delta_{\vec{k}_i, \vec{k}_i(\bc^\prime)} }
\end{eqnarray*}

Focus on the difficult term, introducing the measure $w(c| \bar{k}) = \prod_{i<j}(\frac{\bar{k}}{N} \delta_{c_{ij},1} + (1-\frac{\bar{k}}{N})\delta_{c_{ij},0})$

\begin{eqnarray*}
I = 
\frac
{\sum_{\bc} c_{rs}\prod_{i<j}(\frac{\bar{k}}{N} \delta_{c_{ij},1} + (1-\frac{\bar{k}}{N})\delta_{c_{ij},0})
\prod \delta_{\vec{k}_i, \vec{k}_i(\bc)}
}
{
\sum_{\bc} \prod_{i<j}(\frac{\bar{k}}{N} \delta_{c_{ij},1} + (1-\frac{\bar{k}}{N})\delta_{c_{ij},0})
\prod \delta_{\vec{k}_i, \vec{k}_i(\bc)}
}
\end{eqnarray*}
but I observe that I've already done this calculation. 
\begin{eqnarray*}
I = 
\frac
{
\int d \bomega \rme^{\rmi \bomega \cdot \bk} L(\bomega) 
	\frac
	{	\frac{\bar{k}}{N} 
			\rme^
			{
				-\rmi (\omega_r + \omega_s + \phi_r k_s+ \phi_s  k_r ) 
			}
	}
	{
		\frac{\bar{k}}{N} 
			\left(
			\rme^
			{		
				-\rmi (\omega_r + \omega_s + \phi_r k_s+ \phi_s  k_r ) 
			}
			-1
			\right)
			+1
	}
}
{
\int d \bomega \rme^{\rmi \bomega \cdot \bk} L(\bomega)
}
\end{eqnarray*}
with $L(\bomega)$ defined as before. Since we are talking the large $N$ limit, we can write this as
\begin{eqnarray*}
I = 
\frac{\bar{k}}{N} 
(1+ \order(N^{-1}))
\frac
{
\int d \bomega \rme^{\rmi \bomega \cdot \bk} L(\bomega) 
			\rme^
			{
				-\rmi (\omega_r + \omega_s + \phi_r k_s+ \phi_s  k_r ) 
			}
}
{
\int d \bomega \rme^{\rmi \bomega \cdot \bk} L(\bomega)
}
\end{eqnarray*}

and follow the previous line of reasoning in order to achieve the form 

\begin{eqnarray*}
I = 
\frac{\bar{k}}{N} 
(1+ \order(N^{-1}))
\frac
{
\int \rmd \bomega  \{ \rmd P \rmd \hat{P} \} \rme^{N \Psi[P, \hat{P}]}
			\frac		
			{	
				\int \rmd \bomega		
				\rme^
				{
					\rmi
					\left[
						 \omega \left( k_r -1 \right) + 
						 	\phi \left(m_r-k_s \right)
						 	- \hat{P}\left( \omega, k_r \right)
					\right]
				}
			}
			{	
				\int \rmd \bomega		
				\rme^
				{
					\rmi
					\left[
						 \omega k_r  
						 +	\phi m_r
						 	- \hat{P}\left( \omega, k_r \right)
					\right]
				}
			}
			\frac		
			{	
				\int \rmd \bomega		
				\rme^
				{
					\rmi
					\left[
						 \omega \left( k_s -1 \right) + 
						 	\phi \left(m_s-k_r \right)
						 	- \hat{P}\left( \omega, k_s \right)
					\right]
				}
			}
			{	
				\int \rmd \bomega		
				\rme^
				{
					\rmi
					\left[
						 \omega k_s  
						 +	\phi m_s
						 	- \hat{P}\left( \omega, k_s \right)
					\right]
				}
			}
}
{
\int \rmd \bomega  \{ \rmd P \rmd \hat{P} \}\rme^{N \Psi[P, \hat{P}]}
}
\end{eqnarray*}

We can now proceed with the same substitution that was previously helpful: $\rme^{-\rmi \hat{P}} = Q(\bomega, \bk) = \exp( \bar{k} \sum_{\xi \in \mathbb{N}^2} \gamma(k, \xi) \rme^{-\rmi \bomega \cdot \bxi})$. The integral $\int \rmd \bomega Q(\bomega, \bk) \rme^{\rmi(\omega k_r + \phi m_r)}$ evaluates by series expansion as
$$\int \rmd \bomega Q(\bomega, \bk) \rme^{\rmi(\omega k_r + \phi m_r)} = 
\sum_{\{\xi_1...\xi_{k_r} \}} \prod \gamma(k_r, \xi_s) \delta_{m_r, \sum \xi_s} \frac{\bar{k}^{k_r}}{k_r !}$$

Hence, in leading order, the integral evaluates to 
$$
I = \frac{\bar{k}}{N} \frac{k k^\prime}{\bar{k}^2}
\frac
{
\sum_{\{\xi_1...\xi_{k-1} \}} \prod_{s=1}^{k-1} \gamma(k, \xi_s) \delta_{m-k^\prime, \sum \xi_s}}
{\sum_{\{\xi_1...\xi_{k} \} } \prod_{s=1}^{k} \gamma(k, \xi_s) \delta_{m, \sum \xi_s} }
\frac
{
\sum_{\{\xi_1...\xi_{k^\prime-1} \}} \prod_{s=1}^{k^\prime-1} \gamma(k^\prime, \xi_s) \delta_{m^\prime-k, \sum \xi_s}}
{\sum_{\{\xi_1...\xi_{k^\prime} \} } \prod_{s=1}^{k^\prime} \gamma(k^\prime, \xi_s) \delta_{m^\prime, \sum \xi_s} }
$$

\begin{eqnarray*}
W(\vec{k}, \vec{k^\prime})
&=&
\frac{k k^\prime}{(\bar{k}N)^2}
\sum_{rs} 
\sum_{\{ \vec{k}_1...\vec{k}_N \}}
\prod p(\vec{k}_i)
\delta_{\vec{k}, \vec{k}_r} 
\delta_{\vec{k}^\prime, \vec{k}_s} 
\gamma^\star(\vec{k}, k^\prime)
\gamma^\star(\vec{k^\prime}, k)
\end{eqnarray*}

for $\gamma^\star(\vec{k^\prime}, k) = \frac
{
\sum_{\{\xi_1...\xi_{k^\prime-1} \}} \prod_{s=1}^{k^\prime-1} \gamma(k^\prime, \xi_s) \delta_{m^\prime-k, \sum \xi_s}}
{\sum_{\{\xi_1...\xi_{k^\prime} \} } \prod_{s=1}^{k^\prime} \gamma(k^\prime, \xi_s) \delta_{m^\prime, \sum \xi_s} }$

satisfying $\sum_{m^\prime} \frac{k^\prime}{\bar{k}} p(\vec{k}^\prime)\gamma^\star(\vec{k^\prime}, k) = \gamma(k, k^\prime)$

whereupon we can simplify

\begin{eqnarray*}
W(\vec{k}, \vec{k^\prime})
&=&
\gamma^\star(\vec{k}, k^\prime)
\gamma^\star(\vec{k^\prime}, k)
\frac{k k^\prime p(\vec{k}) p(\vec{k}^\prime)}{(\bar{k}N)^2}
\sum_{rs} 
\sum_{\{ \vec{k}_r, \vec{k}_s \}}
\delta_{\vec{k}, \vec{k}_r} 
\delta_{\vec{k}^\prime, \vec{k}_s}  \\
&=&
\gamma^\star(\vec{k}, k^\prime)
\gamma^\star(\vec{k^\prime}, k)
\frac{k k^\prime p(\vec{k}) p(\vec{k}^\prime)}{(\bar{k})^2}
\end{eqnarray*}

We can very easily take the marginal of the object $W$ using the relation for $\gamma^\star$:

\begin{eqnarray*}
\sum_{m, m^\prime} W(\vec{k}, \vec{k}^\prime)= \gamma(k, k^\prime) \gamma(k^\prime, k)
\end{eqnarray*}

which is in line with our earlier intuition, and serves to validate the preceding steps, e.g. since it shows we are correctly normalised. 

\section{Details of numerical implementation}
\label{sec:algortithm}

There is no subroutine built into the C++ standard template library which calculates a factorial. Even for small values of $x$, the size of $x!$ can become unmanageable for the operating system. A simple factorial subroutine storing values as integers implemented on a 32bit system will fail at about $13!$. Hence an object like ${x \choose y}$ - which is natural and familiar for pen and paper calculations - is not effective when directly programmed according to the definition. 

Additionally, the result found in \ref{eq:answer_gd} has the nuance that where the degree distribution does not have compact support, `impossible' values must be excluded. Fortunately, it is possible to implement an iterative algorithm, which can efficiently calculate the final term, and can be directly adjusted to remove forbidden values. 

Recall that the term which we wish to calculate from \ref{eq:answer_gd} is $\sum_{k,m} p(k,m) \log \mathscr{P}_{k,m}$ where $\mathscr{P}_{k,m}\leq \mathscr{U}_{k,m} = \sum_{\{ \xi_1...\xi_k \}} \delta_{\sum \xi_i , m} \prod_i \left( 1 - \delta_{p(\xi_i),0} \right)$. Observe that $\mathscr{U}_{k,m}$ must satisfy 
\begin{eqnarray}
\label{eq:iterative_relation_for_combinatorial_term}
\mathscr{U}_{k,m} = \sum_s \mathscr{U}_{k-1,m-s} \mathscr{I}(s)
\end{eqnarray}
where $\mathscr{I}(s)$ is an indicator: $\mathscr{I}(s) = 1 - \delta_{p(s),0}$. That is to say, the number of configurations generalised degree $(k,m)$ with a certain degree value $s$ of the first neighbour is exactly the number of configurations with generalised degree $(k-1, m-s)$ (i.e. considering as though the first neighbour was removed, since its degree value is known). Hence, the total number of configurations is calculated by summing over all values of $s$. 

We set up an iterative scheme as follows: populate row 1 with 0 or 1 depending if the degree i has a zero or non-zero probability. Populate row 2 as 1, 2, 3, 4.... However, wherever row 1 has a zero value, reduce the corresponding entry in row 2 by 1 instead of incrementing. This row can be interpreted as follows: the $i$th entry corresponds to the number of possible partitions of second degree $i-1$ between 2 direct neighbours. This is consistent with \ref{eq:iterative_relation_for_combinatorial_term}. Populate row 3 based on \ref{eq:iterative_relation_for_combinatorial_term}. To populate row 4, we only need rows 1 and 3. Hence, we can release some memory by evaluating $\sum_{k=1,2;m} p(k,m) \log \mathscr{U}_{k,m}$, storing this value, and releasing (i.e. over-writing) row 2. Note that the value of $\mathscr{U}_{k,m}$ is read from the entry in row $k$ entry $m-k+1$ . This scheme allows full accuracy to be maintained to much higher values of $k$ and $m$ - and allows direct adjustment for the case of `impossible' degree values. 

\section{Validation}
\label{sec:gd_synthetic_examples}

The entropy expression must satisfy certain properties. We can use these to check if our claimed answer (equation \eref{eq:entropy_final}) is plausible. 
Where there is a simple enough example to be enumerated directly, we expect to be able to reconcile the analytical expression for $Z$ with counting $\sum_c \delta_{\vec{k}, \vec{k(c)}}$
Where the self consistency relation for $\gamma$ can be solved directly, we expect to be able to reconcile the two forms of the $\Gamma$ (the one involving $\gamma$ terms, and the one involving $W$ terms)
We expect to retrieve the old simple degree result if we insert a `flat' generalised degree distribution. 

\subsection{Ladder configuration}
\label{sec:ladder}
\begin{figure}
\begin{minipage}[b]{0.45\linewidth}
\includegraphics[scale=0.2,  angle = 90]{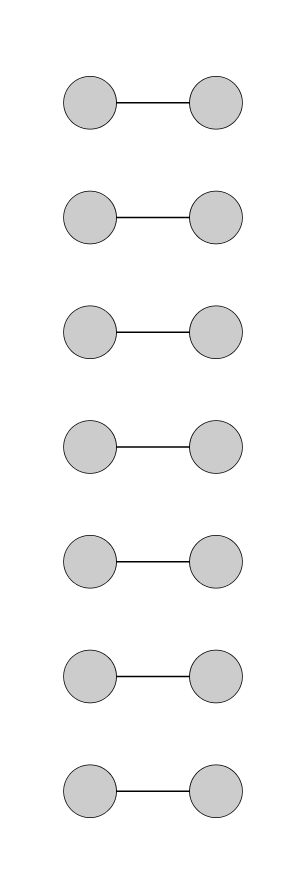} 
\end{minipage}
\hspace{0.2\linewidth}
\begin{minipage}[t]{0.25\linewidth}
  \vspace*{-40pt}
\begin{tabular}{| l| c| }
\hline
   $k$ & 1  \\
   $m$ & 1  \\
   \hline
   $p(k,m)$ & 1 \\
\hline
 \end{tabular}
 \end{minipage}
\caption{A validation example with only one degree value - the ladder. In \ref{sec:ladder} the partition function is evaluated directly and compared to the result using equation \eref{eq:entropy_final}.}
\label{fig:ladder}
\end{figure}
For this case (illustrated in figure \ref{fig:ladder}), we find that $\gamma(1,1)= 1$,$W(1,1)= 1$ and $\mathscr{P}= 1$, so  $\Gamma = 0$ calculated either way.              
To calculate $S$ from first principles for this network we need to calculate $p(\bc) = \frac{\prod_i p(k_i)}{ Z(\bk_i)}$
 where we have defined our ensemble as drawing degrees from $p(k)= \delta_{k,1}$ . As always, the core of the problem is in calculating 
 $Z(\bk)= \sum_c \prod_i \delta_{1, k_i(c)}$ which is equivalent to counting the permutations of the network drawn above. There are $N!$ orderings of the node, but this has to be divided by $2$ since the bonds are symmetric and by $\frac{N}{2}$ since the order of the bonds is immaterial. 
$$\sum_{\bc} p(\bc) \log(p(\bc)) = - \log\frac{N!}{2 \frac{N}{2}!} = - \log N! + log 2 + \log \frac{N}{2}!$$ 
which by Stirling's approximation goes to
$  = \frac{-N}{2} \left( \log{N} -1 \right) $
which corresponds to our analytical result in leading order in $N$.
\subsection{Wheel configuration}
\label{sec:wheel}
\begin{figure}
\begin{minipage}[b]{0.45\linewidth}
\includegraphics[width = \linewidth]{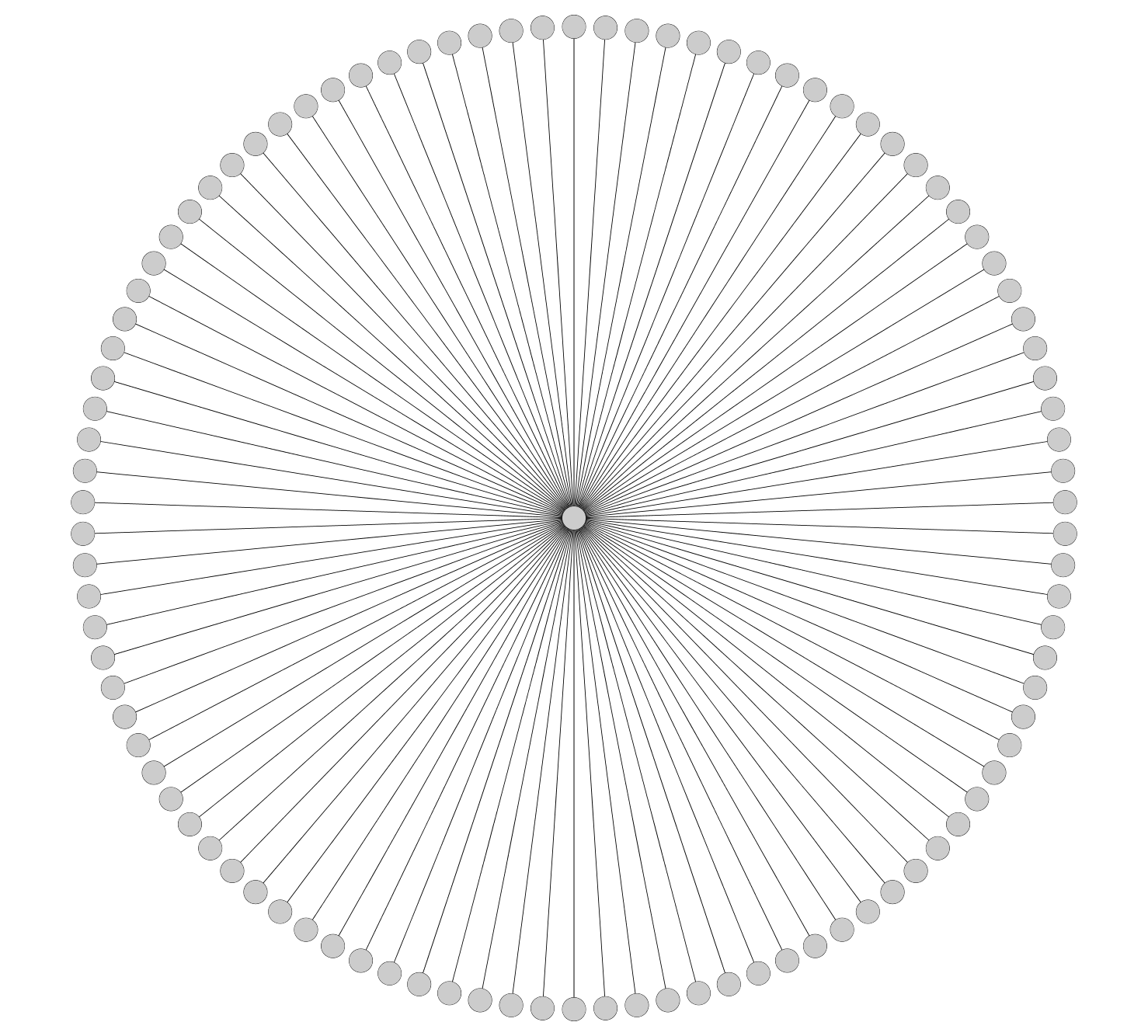}
 \end{minipage}
 \hspace{0.2\linewidth}
\begin{minipage}[t]{0.25\linewidth}
  \vspace*{-80pt}
\begin{tabular}{| l| c| c| }
\hline
   $k$ & 1 & 99 \\
   $m$ & 99 & 99 \\
   \hline
   $p(k,m)$ & 99/100 & 1/100 \\
\hline
 \end{tabular}
 \end{minipage}
\caption{A validation example with two degree values - the wheel. In \ref{sec:wheel} the partition function is evaluated directly and compared to the result using equation \eref{eq:entropy_final}.}
\label{fig:wheel}
\end{figure}
Consider an ensemble with generalised degrees consistent with figure \ref{fig:wheel}. Based on the self consistency equations we find that $\gamma(1,s)\gamma(s,1) = \frac{1}{2}$  and so it follows:
\begin{eqnarray}
\label{eq:Gamma_wheel} \! \! \! \!\!\!\!\!\!\!\!
\Gamma = \frac{99}{100} \log \gamma(1,s) + \frac{99}{100}\log \gamma(s,1) =  \frac{99}{100} \log \gamma(1,s)\gamma(s,1)= - \frac{\bar{k}}{2}log 2
\end{eqnarray}
We can immediately deduce the key parameters of this problem worked through as a starting point for an ensemble with given generalised degrees. 
\begin{eqnarray*} \bar{k} = 2 \times 99/100 ~~~ p(k,m) = \frac{\delta_{(k,m), (99,99)}}{100} + \frac{99\delta_{(k,m), (1,99)}}{100}
\end{eqnarray*}
Let us work out $Z(\bk)$ for an arbitrary degree sequence drawn from our distribution. For the generalised degree case the only valid wiring is as illustrated: groups of 99 degree 1 nodes around a degree 99 node. To enumerate the ensemble, view the peripheral nodes as consecutively numbered; observe ${\frac{99N}{100}}!$ orderings of such a numbering - of which there are 99! equivalent permutations within each grouping. Hence, claim $Z(\bk)= \frac{(99N/100)!}{99!^{N/100}}$  . 
So, directly calculated 
\begin{eqnarray}
\label{eq:wheel_enumerated}
\frac{1}{N}log Z|_{enumerated} = \frac{99}{100}\left(log(N) + log(99/100) - \frac{1}{100}log 99! \right) - \frac{99}{100} 
\end{eqnarray}
whereas picking up from the appropriate point in the calculation finds this quantity to be
\begin{eqnarray*}\frac{1}{N}log Z|_{analytical} = \frac{\bar{k}}{2}(1+ log(\frac{N}{\bar{k}})) + \sum p(k,m) \log \pi(k) + \Gamma
\end{eqnarray*}
The $\pi(k)$ factors evaluate to 
$$ \log(\pi(1))= \log(\frac{2 \times 99}{100})-\bar{k}$$
$$ \log(\pi(99))= \log\left[\frac{2 \times 99}{100}\right]^{99}- \bar{k}-  \log 99!$$
so, inserting equation \eref{eq:Gamma_wheel} into \ref{eq:wheel_enumerated} shows that
$$\frac{1}{N}log Z|_{analytical} =- \frac{\bar{k}}{2}+ \frac{\bar{k}}{2}log N + \frac{\bar{k}}{2}log \bar{k} - \frac{1}{100}log 99! + \Gamma$$
which reconciles with the solution from first principles, as required.
%

\subsection{A validation example with a more complicated degree-degree correlation}
\label{sec:more_complicated_ddc}
\begin{figure}[ht]
\begin{minipage}[b]{0.45\linewidth}
\includegraphics[width =  0.6\linewidth]{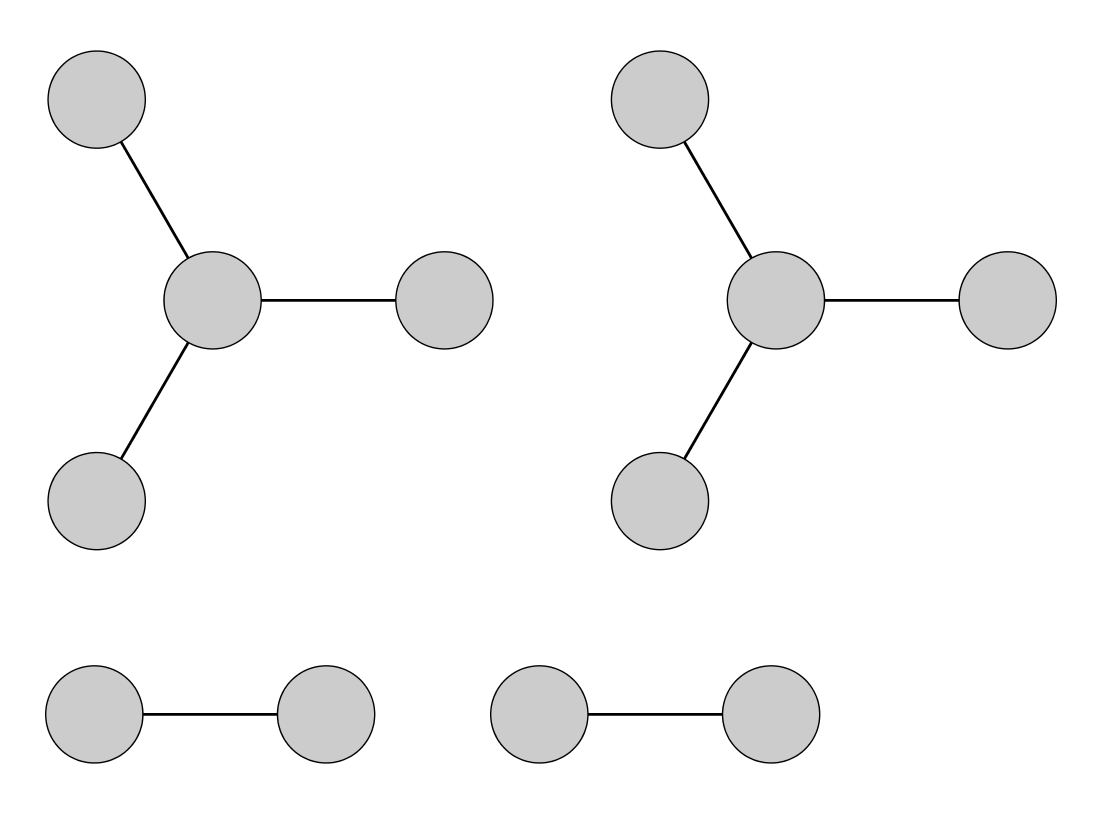}
\vspace{-10pt} 
 \end{minipage}
 \hspace{0.2\linewidth}
\begin{minipage}[t]{0.25\linewidth}
\vspace{-40pt} 
\begin{tabular}{| l| c| c| c|}
\hline
   $k$ & 1 & 1 & 3\\
   $m$ & 1 & 3 & 3 \\
   \hline 
   $p(k,m)$ & 1/3 & 1/2 & 1/6 \\
\hline
 \end{tabular}
 \end{minipage}
\caption{A validation example with non-constant degree-degree correlation. In \ref{sec:more_complicated_ddc} the partition function is evaluated directly and compared to the result using equation \eref{eq:entropy_final}.}
 \label{fig:gd_W_example}
 \end{figure}

Consider a network based on figure \ref{fig:gd_W_example} as a repeating motif. In fact, this is the only possible wiring of such a network that fulfils the generalised degree distribution required, up to permutation of nodes. 

As previously, it is convenient to work with $\gamma^\star(\vec{k^\prime}, k) = \frac
{
\sum_{\{\xi_1...\xi_{k^\prime-1} \}} \prod_{s=1}^{k^\prime-1} \gamma(k^\prime, \xi_s) \delta_{m^\prime-k, \sum \xi_s}}
{\sum_{\{\xi_1...\xi_{k^\prime} \} } \prod_{s=1}^{k^\prime} \gamma(k^\prime, \xi_s) \delta_{m^\prime, \sum \xi_s} }$
satisfying $\sum_{m^\prime} \frac{k^\prime}{\bar{k}} p(\vec{k}^\prime)\gamma^\star(\vec{k^\prime}, k) = \gamma(k, k^\prime)$
%
%

But the example has been constructed so that (up to permutation) only one sequence is graphical: $\{ (1,1), (1,1), (1,3), (1,3), (1,3), (3,3) .... \}$ repeated $N/6$ times. We can do the calculations for the small network only without loss of generality.  

The self consistency relations are trivial in this example
$$ \gamma^2(1,1) = \frac{p(1,1)}{\bar{k}} = \frac{1}{4}$$
$$ \gamma(1,3)\gamma(3,1) = \frac{3 p(3,3)}{\bar{k}}= \frac{3}{8}$$
from which it follows that $\Gamma$ evaluated from equation \eref{eq:intermediate_form} is
\begin{eqnarray*}
\Gamma & = & \frac{\log \gamma(1,1)}{3} + \frac{\log \gamma(1,3)}{2} + \frac{\log \gamma^3(3,1)}{6} \\
&=& \frac{1}{6} \log \left[ \frac{1}{4} \left(\frac{3}{8} \right)^3\right]
 \end{eqnarray*}
Which evaluates to the same as the expression for $\Gamma$ proposed in equation \eref{eq:entropy_final}
\begin{eqnarray}
\label{eq:gamma_fram_validation_example_with_two_W_values}
 \Gamma &=& \frac{\bar{k}}{2} \sum_{k, k^\prime} W(k, k^\prime) \log W(k, k^\prime) + \sum_{k,m} p(k,m) \log {m-1 \choose k-1} \\ \nonumber
 &=& \frac{8}{6} \frac{1}{2} \left[ \frac{1}{4} \log \frac{1}{4} + \frac{3}{8} \log \frac{3}{8} + \frac{3}{8} \log \frac{3}{8} \right] + 0 \\ \nonumber
 &=& \frac{1}{6} \log \left[ \frac{1}{4} \left(\frac{3}{8} \right)^3\right] 
\end{eqnarray}
as predicted. 

A combinatorial argument can be used to calculate $Z(\bk)$ in this case. Consider the network to be a combination of two sub networks. The sub-network of nodes degree $(1,1)$ has $N/3$ nodes. Repeat the earlier strategy of counting the configurations by counting the number of labellings of the diagram ($\frac{N}{3}!$), and then dividing by the symmetries: $2!^{N/6}$ since the pairs are symmetric and $(N/6)!$ since the order doesn't matter. In the other subnetwork, there are $(N/2)!$ orderings of the peripheral nodes, divided by $(3!)^{N/6}$ symmetry factor (but the 'order' does matter, since this fixes the central node). Hence claim
$$ Z(\bk)= \frac{\frac{N}{3}! \frac{N}{2}!}{\frac{N}{6}! 2!^{N/6}3!^{N/6}}$$
If $N=6$, then $Z(\bk) = \frac{2!3!}{1! 2! 3!}=1$ as expected. If $N=12$, then $Z(\bk) = \frac{4!6!}{2! 2!2! 3!3!} = \frac{6!}{ 2!3!}= 60$ which can be corroborated directly. 

Working from first principles, it follows that
\begin{eqnarray*} \frac{1}{N} \log Z(\bk) &=& \frac{1}{3}\log\left(\frac{N}{3} \right)+ \frac{1}{2}\log\left(\frac{N}{2} \right)-\frac{1}{6}\log\left(\frac{N}{6}\right)
 -\frac{1}{6}\log\left(12 \right) - \frac{4}{6}\\
&=& \frac{4 }{6} \log N -\frac{1}{3}\left[ \log 3 + 2 \log 2 \right]  - \frac{2}{3}
\end{eqnarray*}
%
This matches the analytical result, evaluated with the help of equation \eref{eq:gamma_fram_validation_example_with_two_W_values}.
\subsection{A connected network example}
\label{sec:connected_network}
\begin{figure}[ht]
\begin{minipage}[b]{0.45\linewidth}
\includegraphics[width = 0.6\linewidth]{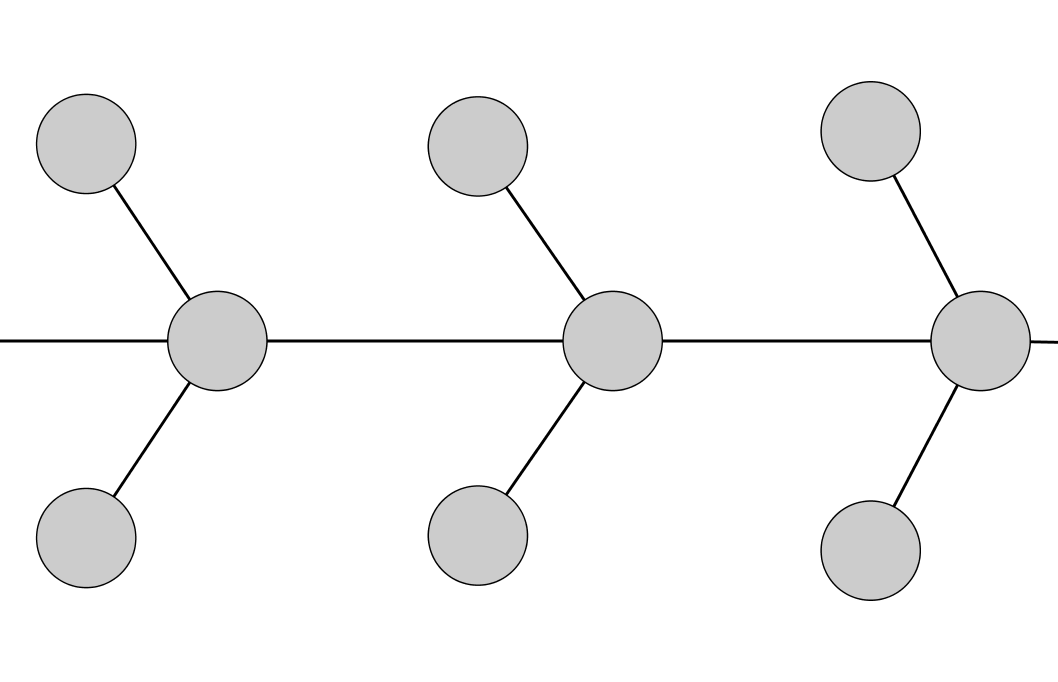}
 \end{minipage}
  \hspace{0.2\linewidth}
\begin{minipage}[t]{0.25\linewidth}
  \vspace*{-40pt}
\begin{tabular}{| l| c| c| c|}
\hline
   $k$ & 1 & 4 \\
   $m$ & 4 & 8 \\
   \hline
   $p(k,m)$ & 2/3 & 1/3  \\
\hline
 \end{tabular}
 \end{minipage}
\caption{A validation example based on a connected network. In \ref{sec:connected_network} the partition function is evaluated directly and compared to the result using equation \eref{eq:entropy_final}.}
\label{fig:gd_molecule}
 \end{figure}
Consider the ensemble of networks defined by the general degree sequence set out in figure \ref{fig:gd_molecule}. By combinatorics, argue that there are $\frac{N}{3}!$ orderings of the centre nodes, and $\frac{2N}{3}!$ orderings of the remaining nodes, divided by $2^{\frac{N}{3}}$ for symmetry. Hence it follows that
\begin{eqnarray*}
\frac{1}{N} \log Z(\bk, \bm)
 &= \frac{1}{N} 
 \log 
 \left( 
 		\frac{
 				\frac{N}{3}!
 				\frac{2N}{3}!
 			}
 			{
 				2^{\frac{N}{3}}
 			}
 \right) 
 \\
&= \frac{1}{3} 
\log 
 \frac{N}{3} 
 +  
 \frac{2}{3} 
 \log  
 \frac{2N}{3} -  
 \frac{1}{3} \log  2  -1
 \\
&= \log N - \log 3 + \frac{1}{3} \log 2 -1
\end{eqnarray*}
Working using the analytical formula, note immediately that $\mathscr{P}_{k,m} = 1$ for all $k,m$ appearing in the network (by construction), and $\frac{\bar{k}}{2}=1$ so
\begin{eqnarray*}
\frac{1}{N} \log Z(\bk, \bm) &=\log N + \log2 -1 -\frac{2}{3} \log 2 - \log 3 \\
&= log N + \frac{1}{3} \log 2 - \log 3 -1
\end{eqnarray*}

\end{document}